\documentclass[twocolumn,preprintnumbers,amsmath,amssymb]{revtex4-2}
\usepackage{comment} 
\usepackage{graphicx}
\usepackage{dcolumn}
\usepackage{bm}
\usepackage[compact]{titlesec}

\usepackage{slashed}

\usepackage{xcolor}

\usepackage{hyperref}
\renewcommand{\arraystretch}{1.8}       
\newcommand{\Matter}{\scriptstyle{\rm{Matter}}}
\newcommand{\gD}{{\mathbf{D}}}

\newcommand{\YM}{\scriptstyle{\rm{YM}}}

\newcommand{\particle}{\scriptstyle{\rm{particle}}}
\newcommand{\half}{{{\textstyle\frac{1}{2}}}}
\newcommand{\quarter}{{{\textstyle\frac{1}{4}}}}
\newcommand{\be}{\begin{equation}}
\newcommand{\ee}{\end{equation} }
\newcommand{\beqa}{\begin{eqnarray} }
\newcommand{\eeqa}{\end{eqnarray} }
\newcommand{\ba}{\begin{array}}
\newcommand{\ea}{\end{array}}
\newcommand{\bpm}{\begin{pmatrix}}
\newcommand{\epm}{\end{pmatrix}}

\newcommand{\so}{\mathbf{so}}

\newcommand{\GL}{\mathbf{GL}}
\newcommand{\dis}{\displaystyle}

\newcommand{\rmd}{{\rm d}}
\newcommand{\rd}{{\rmd}}

\newcommand{\rmD}{{\rm D}}

\newcommand{\ODD}{\mathbf{O}(D,D)}

\newcommand{\brpartial}{\bar{\partial}}

\newcommand\Tr{{\rm Tr}}

\newcommand\whf{\widehat{f}}
\def\wtf{\widetilde{f}}
\def\wtu{\widetilde{u}}

\newcommand\cB{{\cal B}}

\newcommand\cD{{\cal D}}

\newcommand\cF{{\cal F}}

\newcommand\cH{{\cal H}}

\newcommand\cJ{{\cal J}}

\newcommand\cL{{\cal L}}

\newcommand\cV{{\cal V}}

\newcommand\brcD{\bar{\cD}}

\newcommand\hcL{{\hat{\cal L}}}

\newcommand\fD{{\mathfrak{D}}}

\newcommand\mcH{\mathring{\cH}}
\newcommand\mcV{\mathring{\cV}}

\def\tx{\tilde{x}}
\def\ty{\tilde{y}}

\def\tJ{\tilde{J}}

\def\tpartial{\tilde{\partial}}

\def\na{\nabla}

\def\brrho{\bar{\rho}}
\def\brxi{\bar{\xi}}
\def\brzeta{\bar{\zeta}}
\def\brlambda{\bar{\lambda}}

\def\bri{\bar{\imath}}
\def\brj{\bar{\jmath}}

\def\brn{{\bar{n}}}
\def\brp{{\bar{p}}}

\def\bry{\bar{y}}

\def\brUpsilon{{{\bar{\Upsilon}}}}

\def\brP{\bar{P}}

\def\brV{\bar{V}}

\def\brX{\bar{X}}
\def\brY{\bar{Y}}

\def\brLambda{\bar{\Lambda}}

\newcommand\mP{\mathring{P}}
\newcommand\mbrP{\mathring{\brP}}

\newcommand{\deltaM}{\delta_{\scriptscriptstyle{\rm{M}}}}

\newcommand\vJ{\mathbb{J}} 

\def\aa{\mathtt{a}}
\def\bb{\mathtt{b}}

\def\B{\mathtt{B}}

\begin{document}

\preprint{APCTP Pre2021-032}
\title{Fractons, non-Riemannian Geometry, and Double Field Theory}

\author{Stephen Angus}
\email{stephen.angus@apctp.org}
\affiliation{Asia Pacific Center for Theoretical Physics, Postech, Pohang 37673, Korea}

\author{Minkyoo Kim}
\email{mkim@sogang.ac.kr}
\affiliation{Center for Quantum Spacetime, Sogang University, 35 Baekbeom-ro, Mapo-gu,  Seoul  04107, Korea}

\author{Jeong-Hyuck Park}
\email{park@sogang.ac.kr}
\affiliation{Department of Physics, Sogang University, 35 Baekbeom-ro, Mapo-gu, Seoul 04107,  Korea}

\begin{abstract}
\noindent  
We initiate a systematic study of fracton physics within the geometric framework of Double Field Theory. We ascribe the immobility and large degeneracy of the former to the non-Riemannian backgrounds of the latter, in terms of  generalised geodesics and infinite-dimensional isometries.   A doubled pure Yang--Mills or Maxwell theory reduces to an ordinary  one coupled to a strain tensor of elasticity theory, and thus rather remarkably  provides    a unifying  description of  photons and phonons. Upon a general  Double Field Theory background, which consists of Riemannian and non-Riemannian subspaces,  the dual photon-phonon pair  becomes fractonic over the non-Riemannian subspace.  When  the   elasticity displacement vector condenses, minimally coupled charged particles acquire an effective mass even in the purely Riemannian case, yielding predictions for polaron physics and time crystals.  Furthermore, the immobility of neutral particles along the non-Riemannian directions is lifted to a saturation velocity for charged particles.  
Utilising the differential geometry of  Double Field Theory   we also present curved spacetime extensions which exhibit  general covariance.

\end{abstract}

                             
\maketitle

\section{Introduction\label{SECI}}
Fractons are novel quasiparticles
with properties that challenge the conventional understanding of topological phases
of matter~\cite{Chamon:2004lew,Haah:2011drr, Vijay:2015mka, Vijay:2016phm}.  
In  modern condensed matter physics, it is  generically expected  that any lattice model with local interactions  admits a well-defined continuum field theory limit  in the far infrared regime. However,  fractons defy this doctrine and appear to  require  ingenious or  exotic field theories with some manner of  UV/IR mixing~\cite{Seiberg:2019vrp, Seiberg:2020bhn, Seiberg:2020wsg, Seiberg:2020cxy, Gorantla:2020xap, Rudelius:2020kta, Gorantla:2021svj, Gorantla:2021bda}. 
Fractons have further characteristic properties such as  immobility, infinite ground-state degeneracy in the continuum limit, and higher moment  conservation laws.
We refer readers to \cite{Nandkishore:2018sel, Pretko:2020cko} for reviews and   \cite{Pretko:2016kxt, Pretko:2016lgv, Slagle:2018kqf, Gromov:2018nbv, Pretko:2017fbf, Yan:2018nco, Gromov:2020yoc, Casalbuoni:2021fel, Geng:2021cmq, Qi:2020jrf, Distler:2021qzc, Pretko:2017kvd, Gromov:2017vir, Pretko:2018tit, fractongauge, Pretko:2019omh, Fruchart:2019qnn, Nguyen:2020yve, Choi:2021kmx, Hirono:2021lmd, Bulmash:2018lid, Ma:2018nhd, Slagle:2020ugk, Manoj:2020bcz, Radzihovsky:2020xct, Sun:2021jzs, Yuan:2019geh, Chen:2020jew, Li:2021rga, Distler:2021bop, Yuan:2022mns, Gorantla:2022eem, Yuan:2022wxb, Perez:2022kax} for further significant developments.

Recent advances have shown further that the immobility can be explained in terms of certain  subsystem symmetries  or conserved higher multipole moments.  As symmetry  has been  a successful guiding principle in modern physics,  the characteristics of fracton physics  can be grasped through the underlying (though rather exotic) symmetry laws. For example, a charged particle with both monopole and dipole conservation in  specific directions explains the immobility of the monopole in the corresponding subspace.

Parallel to the endeavours to find continuum field theory limits of all known fracton lattice models, it may be worthwhile to have a novel formalism which allows us to construct systematically and geometrically   new types of quantum field theories   featuring fractons.

In this paper, we  launch  a systematic top-down approach to   fracton physics by employing the geometric framework of Double Field Theory (DFT), {assuming} the $\ODD$ symmetry therein as the first principle.   Historically, ${\mathbf{O}(d,d;\mathbb{Z})}$  was an `emerging' discrete symmetry for string theory compactified on a torus background $T^{d}$~\cite{Giveon:1994fu}. However, from the modern DFT point of view,  string theory itself `knows'  the $\ODD={\mathbf{O}(D,D;\mathbb{R})}$ symmetry  regardless of the  chosen background, with $D$ now denoting the full spacetime dimension. The theory is \textit{ab initio}   `covariant' (rather than invariant)  under $\ODD$ symmetry rotations. Only a specific individual background breaks it spontaneously, either fully or partially, such as ${\mathbf{O}(D,D;\mathbb{R})}\rightarrow {\mathbf{O}(d,d;\mathbb{Z})}$ upon the aforementioned toroidal compactification.

We shall demonstrate in the present paper that fracton physics may arise from such fully $\ODD$ symmetric theories when  the background is non-Riemannian, meaning that an invertible metric $g_{\mu\nu}$ is not defined even locally.  Analogous to General Relativity (GR) which describes physics on Riemannian geometries, the (stringy)  gravitational theory for more general geometries, including both Riemannian and non-Riemannian ones,  is DFT. By embedding fracton physics into  DFT, it becomes readily possible to further address    fermionic extensions, supersymmetrisations,  and curved spacetime generalisations, while likely maintaining consistency with string theory or quantum gravity.

  DFT was originally conceived~\cite{Siegel:1993xq,Siegel:1993th,Hull:2009mi,Hull:2009zb,
Hohm:2010jy,Hohm:2010pp} to make manifest the  hidden  symmetry of $D$-dimensional  supergravity  underlying  the so-called  `Buscher rule'~\cite{Buscher:1987sk,Buscher:1987qj}.   In order to do so, the theory demands the coordinates to be formally doubled, $x^{A}=(\tx_{\mu},x^{\nu})$, $\partial_{A}=(\tpartial^{\mu},\partial_{\nu})$, and redefines  the notion of general covariance: under infinitesimal doubled diffeomorphisms $\delta x^{A}=\xi^{A}(x)$,  a covariant tensor density of weight~$w$ transforms     through a generalised Lie derivative,
\be
\ba{lll}
\hcL_{\xi}T_{A_{1}\cdots A_{n}}&=&\xi^{B}\partial_{B}T_{A_{1}\cdots A_{n}}+w\partial_{B}\xi^{B\,}T_{A_{1}\cdots A_{n}}\\
{}&{}&+
\sum_{j=1}^{n}(\partial_{A_{j}}\xi^{B}-\partial^{B}\xi_{A_{j}})T_{A_{1}\cdots B\cdots A_{n}}\,.
\ea
\label{gLie}
\ee
Here  $A,B=1,2,\cdots, D{+D}$, are $\ODD$ 
indices which are raised and lowered by an $\ODD$ invariant metric,
\be
\cJ_{AB}=\left(\ba{cc}{\mathbf{0}}&{\mathbf{1}}
\\{\mathbf{1}}&{\mathbf{0}}\ea\right)\,.
\ee
Closure of \eqref{gLie} requires imposing the so-called  `section condition', 
\be
\partial_{A}\partial^{A}=0\,,
\label{seccon}
\ee 
which enforces that  the contraction between any pair of derivatives should be trivial.  Decomposing  this as  $\partial_{A}\partial^{A}=\partial_{\mu}\tpartial^{\mu}+\tpartial^{\mu}\partial_{\mu}$, the condition is conveniently solved by switching off any tilde-coordinate dependence,  ${\tpartial^{\mu}=0}$. In this way, the theory is not truly doubled: rather, it packages various component fields into a unifying $\ODD$ multiplet.

{In DFT} the `dilaton' $d$ and the `generalised metric' $\cH_{AB}$ are the two fundamental variables that constitute the gravitational sector, in analogy with the Riemannian metric $g_{\mu\nu}$  in GR. While the former exponentiates to a unit-weight scalar density $e^{-2d}$,  the latter satisfies  its own  defining properties, 
 \be
 \ba{ll}
 \cH_{AB}=\cH_{BA}\,,\quad&\quad \cH_{A}{}^{C}\cH_{B}{}^{D}\cJ_{CD}=\cJ_{AB}\,.
 \ea
 \label{defcH}
 \ee
This implies that $\det\cH_{AB}=\pm1$, hence the generalised metric alone cannot produce any integral measure like $\sqrt{g}$ in GR.  Instead,  combined with the $\ODD$ invariant metric, it generates a pair of   mutually orthogonal projectors, $P_{AB}=\half(J+\cH)_{AB}$ and $\brP_{AB}=\half(J-\cH)_{AB}$, satisfying
\[
\ba{lll}
P_{A}{}^{B}P_{B}{}^{C}=P_{A}{}^{C}\,,\quad&\quad\!\!
\brP_{A}{}^{B}\brP_{B}{}^{C}=\brP_{A}{}^{C}\,,\quad&\quad\!\!
P_{A}{}^{B}\brP_{B}{}^{C}=0\,.
\ea
\]
Parallel to General Relativity (GR),   DFT has its own Christoffel symbols $\Gamma_{ABC}$, scalar/Ricci/Einstein curvatures, \textit{etc.}, all arising from  $\{d,\cH_{AB}\}$~\cite{Jeon:2011cn,Park:2015bza}.  Moreover,   when coupled   to extra `matter'  $\ODD$ symmetrically,  DFT satisfies ``Einstein  equations''~\cite{Angus:2018mep},
\be
G_{AB}=T_{AB}\,,
\ee
which unifies  the equations of motion of    $d$ and $\cH_{AB}$. 
The LHS and RHS satisfy  a Bianchi identity and on-shell conservation, respectively: $\na_{A}G^{AB}=0=\na_{A}T^{AB}$, with  $\na_{A}=\partial_{A}+\Gamma_{A}$.   
The extra matter can be quite generic~\cite{Jeon:2011vx,Jeon:2011sq,Jeon:2012hp,Jeon:2011kp,
Hohm:2011zr,Hohm:2011dv,Jeon:2012kd,Angus:2019bqs,Lescano:2021nju},  such as  point particles~\cite{Ko:2016dxa,Basile:2019pic},   strings~\cite{Lee:2013hma,Park:2016sbw},  and  the Standard Model~\cite{Choi:2015bga}.

In the early days of the development of DFT, the generalised  metric was simply assumed  to be of the form
\be
\cH_{AB}=\left(\!\ba{cc}\cH^{\mu\nu}&\cH^{\mu}{}_{\lambda}\\
\cH_{\kappa}{}^{\nu}&\cH_{\kappa\lambda}\ea\!\right)=\left(\!\ba{cc}g^{\mu\nu}&-g^{\mu\rho}B_{\rho\lambda}\\
B_{\kappa\rho}g^{\rho\nu}&g_{\kappa\lambda}-B_{\kappa\rho}g^{\rho\sigma}B_{\sigma\lambda}\ea\!\right).\label{Rp}
\ee
In this case, the $D$-dimensional diagonal blocks roughly correspond to the inverse metric $g^{\mu\nu}$ and metric $g_{\mu\nu}$, with additional components generated by a skew-symmetric tensor `$B$-field', $B_{\mu\nu}$.  Together with the decomposition of the DFT volume element as $e^{-2d} = e^{-2\phi}\sqrt{-g}$, the resulting fields $\{g_{\mu\nu},B_{\mu\nu},\phi\}$ constitute the  gravitational multiplet in the supergravity theory which arises as the low-energy effective description of the massless modes of the closed string propagating in  Minkowskian spacetime.

However,   this is not the most general parametrisation  of the generalised metric that satisfies   the two defining  conditions~(\ref{defcH}). Surprisingly, it  turns out that DFT describes not only the Riemannian geometries given in (\ref{Rp})  but also non-Riemannian ones where an invertible Riemannian metric cannot be defined even locally~\cite{Lee:2013hma}.  Namely, with respect to the section choice~${\tpartial^{\mu}=0}$, the upper-indexed ${D\times D}$ block matrix $\cH^{\mu\nu}$ can be degenerate.  From the most general solutions to the conditions~(\ref{defcH}),   
{all possible  DFT geometries   have been  classified by two non-negative integers $(n,\brn)$, with $\mathbf{dim}\big(\mathbf{ker\,}\cH^{\mu\nu}\big)=n+\brn$~\cite{Morand:2017fnv}.   Only  those of type $(0,0)$  are   Riemannian,   while  others are intrinsically non-Riemannian.  In particular, the maximally non-Riemannian cases of $(D,0)$ or $(0,D)$ correspond to $\cH_{AB}=\pm\cJ_{AB}$, and thus they are the two perfectly symmetric vacua of DFT, preserving the entire $\ODD$ symmetry with  no moduli~\cite{Cho:2018alk}.    Intriguingly then,  the   Riemannian spacetime (\ref{Rp})  may arise  after the spontaneous breaking of  $\ODD$ symmetry,  which identifies   $g_{\mu\nu}$ and $B_{\mu\nu}$   as the  massless  Nambu--Goldstone bosons~\cite{Berman:2019izh}.  Some  intermediate types of non-Riemannian geometries, such as $(1,1)$, $(D{-1},0)$, \textit{etc.,}~\cite{Ko:2015rha,Morand:2017fnv,Berman:2019izh,Blair:2019qwi} have also been identified as non-relativistic or ultra-relativistic gravities and/or  strings~\cite{Gomis:2000bd,Danielsson:2000gi,Andringa:2012uz,Harmark:2017rpg,Bergshoeff:2014jla,Duval:2014uoa,Bekaert:2015xua}.

Splitting the coordinates into three parts,
\be
\ba{ll}
x^\mu = \big(\, x^a\,,\, y^i\,,\, \bry^{\bri}\,\big)\,,\qquad&\quad
\partial_{\mu}=\big(\,\partial_{a}\,,\,\partial_{i}\,,\,\brpartial_{\bri}\,\big)\,,
\ea
\ee
where 
${1\leq a\leq D{-n}{-\brn}}$, ${1\leq i \leq n}$,  ${1\leq\bri\leq\brn}$,       \textit{a flat} $(n,\brn)$ \textit{background} is given by constant $d$ {and}, with a sub-dimensional (Minkowskian) metric $\eta_{ab}$,~\cite{Blair:2020gng}  
\be
\ba{lll}
\cH^{\mu\nu}{=\eta^{ab}}\delta_{a}^{\mu}\delta_{b}^{\nu}\,,~&
\cH_{\mu\nu} {=\delta_{\mu}^{a}}\delta_{\nu}^{b}\eta_{ab}\,,~&
\cH_{\mu}{}^{\nu}{=\delta_{\mu}^{i}}\delta_{i}^{\nu}-
\delta_{\mu}^{\bri}\delta_{\bri}^{\nu}\,,
\ea
\label{flatH}
\ee
while $\cH^{\mu}{}_{\nu}{=\cH_{\nu}}{}^{\mu}$ and ${\Gamma_{ABC}=0}$, hence ${\na_{A}=\partial_{A}}$.  Here `flat'  means simply being constant: unlike GR it appears that there is no four-indexed ``Riemann curvature'' in DFT~\cite{Jeon:2011cn,Hohm:2011si}.   Nevertheless,   any constant  background of $\{d,\cH_{AB}\}$ solves the vacuum Einstein equations~${G_{AB}=0}$, hence in contrast to GR the constant flat geometries  are  not unique in DFT.   The fact that $\cH^{\mu\nu}$ and $\cH_{\mu\nu}$ are degenerate $D\times D$ matrices with ${{n+\brn}\neq0}$  characterises the non-Riemannianity. 

In the present paper, we will further establish connections between double field theory and fracton physics, for generic $(n,\brn)\neq(0,0)$.  We will identify two main points of contact with known fracton models.  The first is the key idea that mobility restrictions arise naturally from non-Riemannian geometry  \textit{a la} (generalised) geodesics, with infinite-dimensional isometries playing the role of higher-multipole conservation laws, as will be explained in detail in the next section.  In addition, we reveal that the DFT generalisation of Yang--Mills theory, to be discussed in section~\ref{SECYM}, secretly contains an elasticity theory.  Theories of elasticity are known to be related to fracton models via a duality transformation~\cite{Pretko:2017kvd}.  Since the elasticity theory is present even for purely Riemannian geometries, this represents a second, independent link to fractons.

The remainder of this paper   is   organised  as follows. In the next \textbf{section~\ref{SECTKM}},  we present three key motivators  for our  proposal of studying  fractons via DFT.  In \textbf{section~\ref{SECPS}}, as elementary warm-up exercises, we consider  a particle action and   scalar field theory on the  non-Riemannian $(n,\brn)$  constant background and  verify their fractonic behaviours.  We then turn to our main example, doubled pure Yang--Mills theory, in \textbf{section~\ref{SECYM}}. We show that it reduces to  ordinary Yang--Mills coupled to an     elasticity theory of a non-Abelian strain tensor, and we spell out its infinite-dimensional Noether symmetries originating from the non-Riemannian isometries.  We subsequently couple its Abelian version, \textit{i.e.~}doubled Maxwell theory, to charged  particles in \textbf{section~\ref{SECMP}} and study the resulting dynamics. In particular, we observe  that the elasticity displacement vector, once condensed, changes the effective mass of the particle. In  \textbf{section~\ref{SECCurved}} we extend  our results to curved  $(n,\brn)$ backgrounds through the DFT formalism. We conclude with comments  including a connection to polarons in \textbf{section~\ref{SECD}}, and we display some technical formulae in \textbf{Appendix~\ref{SECA}}.

While our primary goal was to explore the fratonic nature of field theories on  non-Riemannian DFT backgrounds, during the investigation of the doubled pure Yang--Mills theory, as well as Maxwell theory and its coupling to point particles, we uncovered some  remarkable  properties genuinely  valid  even for Riemannian backgrounds, or on Riemannian subspaces. One is that the doubled pure Yang--Mills theory~(\ref{YM})  reduces to an ordinary (undoubled)   Yang--Mills theory coupled to  a (non-Abelian) strain tensor theory~(\ref{sYM}), such that  its Abelian version  provides  a unifying description of  photons and phonons.  Furthermore, when minimally coupled to a point particle~(\ref{minimalC}), the particle will acquire an effective mass through the condensation of the displacement vector of phonons~(\ref{effMASS}), suggesting a potential application to polaron physics. \\


\section{Three Key Motivators\label{SECTKM}}
The three key motivators  for our  proposal of studying  fractons via DFT are    \textbf{A.}  {geodesic immobility};  \textbf{B.}  {infinite-dimensional isometries}; and  \textbf{C.}  
{induced  Noether currents}. All of these assume  \textit{non-Riemannian constant  backgrounds}~(\ref{flatH}).
\vspace{5pt}
\subsection{Geodesic Immobility}
The geometric meaning of the  section condition~(\ref{seccon}) advocated  in~\cite{Park:2013mpa}  is   that half of the doubled coordinates, \textit{e.g.}~$\tx_{\mu}$, are actually  gauged  as  $\gD x^{A}=\big(\rmd\tx_{\mu}-\mathbf{a}_{\mu},\rmd x^{\nu}\big)$. This  enables us to define  an  $\ODD$-symmetric, 
\textit{doubled-diffeomorphism-invariant},     proper length~\cite{Park:2017snt} and  consequently a particle  action~\cite{Ko:2016dxa,Basile:2019pic},
\be
S_{\scriptscriptstyle{\particle}}=\dis{\int}\rd\tau~\half e^{-1}\gD_{\tau}x^{A}\gD_{\tau}x^{B}\cH_{AB}-\half em^{2}\,,
\label{particle}
\ee
where $e$  (einbein) and $\mathbf{a}_{\mu}$ in $\gD_{\tau}x^{A}$ are auxiliary variables. After Gaussian integration of the $\mathbf{a}_{a}$'s along the Riemannian directions, the above doubled particle action reduces upon the constant $(n,\brn)$ background~(\ref{flatH})  to an undoubled one,~\cite{Morand:2017fnv} (\textit{c.f.~}\cite{Casalbuoni:2021fel}),
\be
S^{(n,\brn)}_{\particle}=\dis{\int}\rd\tau~\half e^{-1}\dot{x}^{a}\dot{x}_{a}  -\half em^{2}
+\Lambda_{i}\dot{y}^{i}+\brLambda_{\bri}\dot{\bry}^{\bri}\,.
\label{Fracton}
\ee
Here $\Lambda_{i}$ and $\brLambda_{\bri}$  originate from the field redefinitions of the gauge components $\mathbf{a}_{i}$ and $\mathbf{a}_{\bri}$, respectively, 
and crucially play the role of Lagrange multipliers,  enforcing  immobility along the non-Riemannian directions, 
 \be
 \ba{ll}
 \dot{y}^{i}=0\,,\quad&\quad \dot{y}^{\bri}=0\,.
 \ea
 \label{immobility0}
 \ee 
 Similarly, on a string worldsheet~\cite{Lee:2013hma,Park:2016sbw}, $y^{i}$ and $\bry^{\bri}$  become chiral and anti-chiral, respectively~\cite{Morand:2017fnv}.  
\vspace{5pt}
\subsection{Infinite-dimensional Isometries}
Our second observation  is   that  the isometry of  the  $(n,\brn)\neq(0,0)$  non-Riemannian constant   background~(\ref{flatH}) is infinite-dimensional~\cite{Blair:2020gng}: the  most general solution to the twofold Killing equations,  
\be
\ba{ll}
{\hcL_{\xi}\cH_{AB}=0}\,,\quad&\quad{\hcL_{\xi}e^{-2d}=0}\,,
\ea \label{KillingDFT}
\ee
is, with $\xi^{A}=(\lambda_{\mu},\xi^{\nu})$, 
\be
\ba{ll}
\xi^{a}=w^{a}{}_{b}x^{b}+\zeta^{a}(y)+\brzeta^{a}(\bry)\,,\quad&\quad
\lambda_{a}=\zeta_{a}(y)-\brzeta_{a}(\bry)\,,\\

\xi^{i}=\omega\brn y^{i}+
\zeta^{i}(y)\,,\quad&\quad\lambda_{i}=\rho_{i}(y)\,,\\
\brxi^{\bri}=-\omega n\bry^{\bri}+\brzeta^{\bri}(\bry)\,,\quad&\quad\brlambda_{\bri}=\brrho_{\bri}(\bry)\,.
\ea
\label{KillingSOL}
\ee
Here  $w_{ab}=-w_{ba}$ (Lorentz symmetry)  and $\omega$ are constants.  All  other parameters  are arbitrary functions of the non-Riemannian coordinates $y^{i}$ or $\bry^{\bri}$,  as displayed in (\ref{KillingSOL}). Furthermore, $\zeta^{i}(y)$ and $\brzeta^{\bri}(\bry)$ should be  divergenceless, 
\be
\ba{ll}
\partial_{i}\zeta^{i}(y)=0\,,\quad&\quad
\brpartial_{\bri}\brzeta^{\bri}(\bry)=0\,,
\ea \label{zetadivconstraint}
\ee
which  ensures that ${\partial_{\mu}\xi^{\mu}=0}$, a requirement following from the Killing equation of the dilaton $d$ (\ref{KillingDFT}).
\vspace{5pt}
\subsection{Induced  Noether Currents}
The third point of interest  relates to 
the  energy-momentum tensor in DFT~\cite{Angus:2018mep},
\be
T^{AB}=-e^{2d}\left[8P^{[A}{}_{C}\brP^{B]}{}_{D}\frac{\delta S_{\Matter}}{\delta\cH_{CD}}+\half \cJ^{AB}\frac{\delta S_{\Matter}}{\delta d}\right]\,.
\label{GeneralT}
\ee
By construction, for arbitrary $\xi^{A}$, it satisfies the off-shell relation
\be
\ba{lll}
\partial_{A}({e^{-2d}T^{A}{}_{B}\xi^{B}})&=&e^{-2d}\xi^{B}\na_{A}T^{A}{}_{B}+\frac{1}{2D}T^{A}{}_{A}\hcL_{\xi}e^{-2d}\\
{}&{}&-\half e^{-2d} (PT\brP)^{AB}\hcL_{\xi}\cH_{AB}\,.
\ea
\label{Tconserv}
\ee
Thus,    for the constant  background~(\ref{flatH}) with the Killing vector~(\ref{KillingSOL})  we acquire an on-shell  conserved   current,
\be
\ba{ll}
\vJ^{\mu}=T^{\mu}{}_{A}\xi^{A}=T^{\mu}{}_{\nu}\xi^{\nu}+T^{\mu\nu}\lambda_{\nu}\,,\quad&\quad \partial_{\mu}\vJ^{\mu}=0\,,
\ea
\label{vJ}
\ee
where we have decomposed $T^{\mu}{}_{A}=(T^{\mu\nu}, T^{\mu}{}_{\nu})$.  Note that there is no special relation between the independent energy-momentum tensor components $T^{\mu\nu}$ and $T^{\mu}{}_{\rho}$: in particular, $T^{\mu\nu}\neq T^{\mu}{}_{\rho}g^{\rho\nu}$, not to mention the absence of an invertible metric $g_{\mu\nu}$ in non-Riemannian geometry.
Explicitly, as a collection of independent currents,
\be
\ba{lll}
\vJ^{\mu}&=&
(T^{\mu}{}_{a}{+\eta_{ab}T^{\mu b}})\zeta^{a}(y)+
(T^{\mu}{}_{a}{-\eta_{ab}T^{\mu b}})\brzeta^{a}(\bry)\\
{}&{}&+\omega(\brn T^{\mu}{}_{i}y^{i}{-n T^{\mu}{}_{\bri}\bry^{\bri}})+
T^{\mu}{}_{i}
\zeta^{i}(y)+T^{\mu}{}_{\bri}\brzeta^{\bri}(\bry)\\
{}&{}&+T^{\mu i}\rho_{i}(y)+T^{\mu \bri}\brrho_{\bri}(\bry)+T^{\mu}{}_{a}w^{a}{}_{b}x^{b}\,.
\ea
\label{superJ}
\ee
Evidently, power series expansions of the local parameters in the coordinates $y^{i}$ and $\bry^{\bri}$ generate infinitely many higher multipole conservation laws. This includes dipole conservation laws generated by the parameter $\omega$ and 
other linear terms from $\big{\{}\zeta^{a}(y), \brzeta^{a}(\bry), \zeta^{i}(y),  \brzeta^{\bri}(\bry)\big{\}}$, modulo $\so(n)\oplus \so(\brn)$ rotations.  Among them, the $(D - n - \brn)(n + \brn)$ linear terms of $\zeta^{a}(y)$ and $\brzeta^{a}(\bry)$ correspond to conventional dipole conservations in the non-Riemannian directions, arising from isometries along the Riemannian subspace.  Meanwhile, the linear terms in $\zeta^{i}(y)$ and $\brzeta^{\bri}(\bry)$ generate further non-Riemannian dipole symmetries.  In all, mobility is restricted in the $(n+\brn)$ non-Riemannian directions. 
In the special cases where ${n=1}$ or ${\brn=1}$, the divergenceless condition~(\ref{zetadivconstraint}) actually  enforces $\zeta^{i}$ or $\brzeta^{\bri}$ simply to be constant, which implies the absence of all higher multipole conservation laws along the non-Riemannian directions.  In particular, $(1,1)$ allows  only    dipole  conservation, corresponding to the finite scale transformation 
\be
\ba{ll}
y~\rightarrow~e^{\omega}y\,,\quad&\quad
\bry~\rightarrow~ e^{-\omega}\bry\,,
\ea
\label{pairSCALE}
\ee
where the two non-Riemannian directions are inversely related.  Note that    the  symmetry is still `supertranslational'  in the Riemannian directions for any $(n,\brn)\neq(0,0)$, as $\zeta^{a}(y)$ and $\brzeta^{a}(\bry)$ appearing  in (\ref{KillingSOL}) are arbitrary functions.

Meanwhile, for the global translational symmetries generated by the constant terms in $\xi^{\mu}$ and $\lambda_{\nu}$, the conservation of the current~(\ref{vJ}) reduces to that of   the  energy-momentum tensors,  
\be
{\partial_{\mu}T^{\mu}{}_{\nu}=0}\,,
\ee
for the  untilde $x^{\mu}$-directions,  and further, inequivalently, 
\be
{\partial_{\mu}T^{\mu\nu}=0}\,,
\ee
for the   tilde $\tx_{\mu}$-directions. The latter can be nontrivial   even after switching off the tilde coordinates,  \textit{i.e.~}setting ${\tpartial^{\mu}=0}$ as our choice of section: as we shall see later,  a scalar field theory has trivial $T^{\mu\nu}$~(\ref{TTscalar}), whereas it is nontrivial for Yang--Mills theory~(\ref{TTYM}). \\

The  three  points  \textbf{A,B,C}  imply that any (double field theorisable) field theory should feature the fractonic properties of higher multipole conservation~\cite{Gromov:2018nbv,Cvetkovic}  and  a huge degeneracy of  quantum states, as there are  infinitely many conserved quantities.  
Intriguingly, the $(1,1)$ non-Riemannian background, corresponding to the non-relativistic string~\cite{Gomis:2000bd}, allows only  dipole conservation  along the pair of non-Riemannian directions, $y,\bry$~(\ref{pairSCALE}) which alludes to UV/IR mixing of these two directions.  This property  is comparable to   known  fracton field theory models~\cite{Pretko:2016kxt,Pretko:2016lgv,Seiberg:2020bhn,Seiberg:2020wsg}.   We stress that all of these are direct consequences of the underlying constant non-Riemannian background.   In the following we verify these  properties explicitly for several examples, such as  particles, scalar fields, doubled Yang--Mills theory, and a strain-Maxwell theory minimally coupled to charged  particles. 
The advantage of embedding fracton physics into DFT is that generalisations to curved geometries,  supersymmetry~\`{a} la \cite{Jeon:2012hp},  and consistent string backgrounds are readily available, by setting $D=10$ or $26$ and $n=\brn$~\cite{Park:2020ixf}.  \\

\section{Particle and Scalar  Field\label{SECPS}}
The doubled energy-momentum tensor of the point particle~(\ref{Fracton})  was obtained   in \cite{Angus:2018mep} from the variation of the covariant particle action~(\ref{particle}) following the prescription~(\ref{GeneralT}), 
\be
\ba{ll}
{T^{\mu\nu}=0}\,,\quad&\quad
T^{\mu}{}_{\nu}(x)=\dis{\int}\rd\tau~\dot{x}^{\mu}(\tau)p_{\nu}\delta^{D}\big(x-x(\tau)\big)\,,
\ea
\ee
where the  delta function  is  defined for the  untilde coordinates~$x^{\mu}-x^{\mu}(\tau)$, and $p_{\mu}=(e^{-1}\dot{x}_{a},\Lambda_{i},\brLambda_{\bri})$ is  the conjugate momentum of $x^{\mu}$, of which all components are constant on-shell. Thus,  conservation indeed holds,
\be
\partial_{\mu}T^{\mu}{}_{\nu}=-\dis{\int}\rd\tau~p_{\nu}\frac{\rd~}{\rd\tau}\delta^{D}\big(x-x(\tau)\big)=0\,.
\label{identicalCONSERVATION}
\ee
Further, from  the on-shell relations
\be
\ba{ll}
T^{a}{}_{c}\eta^{cb}=T^{b}{}_{c}\eta^{ca}\,,\quad&\quad
T^{i}{}_{\nu}=0=T^{\bri}{}_{\nu}\,,
\ea
\label{particleT}
\ee  
the conservation of the  current~(\ref{superJ}) readily follows. The corresponding Noether symmetries  of the reduced particle action~(\ref{Fracton})  inherited from the doubled particle action~(\ref{particle}) read, with (\ref{KillingSOL}),
\be
\ba{llll}
\delta x^{a}=\xi^{a}\,,\quad&\quad\delta y^{i}=\xi^{i}\,,\quad&\quad\delta\bry^{\bri}=\brxi^{\bri}\,,\quad&\quad\delta e=0\,,\\
\multicolumn{4}{c}{\delta\Lambda_{i}=-e\dot{x}^{a}\partial_{i}\zeta_{a}(y)-\omega\brn\Lambda_{i}-\Lambda_{j}\partial_{i}\zeta^{j}(y)\,,}\\
\multicolumn{4}{c}{\delta\brLambda_{\bri}=-e\dot{x}^{a}\brpartial_{\bri}\brzeta_{a}(\bry)+\omega n\brLambda_{\bri}-\brLambda_{\brj}\brpartial_{\bri}\brzeta^{\brj}(\bry)\,.}
\ea
\ee

As a target-spacetime counterpart to the particle action, we turn to a scalar field theory with   Lagrangian (density) $e^{-2d}L_{\Phi}$,  \textit{c.f.~}\cite{Seiberg:2020bhn, Seiberg:2020wsg},
\be
L_{\Phi}=-\half\cH^{AB}\partial_{A}\Phi\partial_{B}\Phi-V(\Phi)=-\half \eta^{ab}\partial_{a}\Phi\partial_{b}\Phi-V(\Phi)\,.
\label{LPhi}
\ee
The doubled energy-momentum tensor is,  from \cite{Angus:2018mep},
\be
\ba{ll}
{T^{\mu\nu}=0}\,,\quad&\quad
T^{\mu}{}_{\nu}=\delta^{\mu}_{a}\partial^{a}\Phi\partial_{\nu}\Phi+\delta^{\mu}_{~\nu}L_{\Phi}\,,
\ea
\label{TTscalar}
\ee
which is conserved on-shell as
\be
\partial_{\mu}T^{\mu}{}_{\nu}=\big[\partial_{a}\partial^{a}\Phi{-V^{\prime}(\Phi)}\big]\partial_{\nu}\Phi=0\,.
\ee
The infinite-dimensional  Noether symmetries for the Killing vector~(\ref{KillingSOL}) are given simply by $\delta\Phi=\xi^{\mu}\partial_{\mu}\Phi$.  In particular,   when  the scalar theory is free with  a Lagrangian  $L_{\Phi}=\half\Phi\big(\eta^{ab}\partial_{a}\partial_{b}\Phi-m^{2}\Phi\big)$ which vanishes on-shell, its  energy-momentum tensor also satisfies (\ref{particleT}).     Thus, the usual agreement  between a spinless particle and a scalar field generalises to   generic $(n,\brn)$ constant non-Riemannian  backgrounds.    It is also worthwhile to note that  massless scalar fields 
propagate through sub-dimensional Riemannian spacetime only: ${\partial_{a}\partial^{a}\Phi=0}$.\\

\section{Doubled Yang--Mills\label{SECYM}}
Our next example is a doubled generalisation of Yang--Mills theory.  This turns out to be a rich theory in its own right (even for purely Riemannian geometries): in the Abelian case, it reduces to a theory of photons and phonons, and thus may itself be applicable to systems of lattice vibrations interacting with light.  Moreover, this suggests a second pathway linking DFT to fracton physics: established fracton models such as symmetric tensor gauge theories are known to be dual to phonon systems via fracton-elasticity duality~\cite{Pretko:2017kvd}.

For a doubled vector potential $\cV_{A}$, the  fully covariant field strength $(P\cF\brP)_{AB}=P_{A}{}^{C}\brP_{B}{}^{D}\cF_{CD}$  is  projected from the ``semi-covariant'' one~\cite{Jeon:2010rw,Jeon:2011cn},
\be
\cF_{AB}=\na_{A}\cV_{B}-\na_{B}\cV_{A}-i[\cV_{A},\cV_{B}]\,.
\label{cF}
\ee
The doubled pure Yang--Mills Lagrangian  $e^{-2d}L_{{\YM}}$ then takes the form~\cite{Jeon:2011kp}
\be
L_{{{\YM}}}=\Tr\big[P^{AC}\brP^{BD}\cF_{AB}\cF_{CD}\big]\,.
\label{YM}
\ee
With $\cD_{A}=\na_{A}-i[\cV_{A},\,\cdot\,\,]$, the  equations of motion are~\cite{Park:2015bza}
\be
\cD_{A}
(P\cF\brP)^{[AB]}=\half
\cD_{A}\big[
(P\cF\brP)^{AB}+(\brP\cF P)^{AB}\big]=0\,,
\label{EOM}
\ee
while the energy-momentum tensor is~\cite{Angus:2018mep}
\be
\ba{lll}
T_{AB}&=&-4P_{[A}{}^{C}\brP_{B]}{}^{D}\Tr\!\left[(\cF\cH\cF)_{CD}+\cD_{E}\!\left(\cF_{CD}\cV^{E}\right)\right]\\
{}&{}&+\cJ_{AB}L_{\YM}\,.
\ea
\label{YMT}
\ee

We now compute   the Lagrangian explicitly  on the 
constant $(n,\brn)$ non-Riemannian background~(\ref{flatH}), which we denote using $\{H^{\mu\nu},K_{\mu\nu},Z_{\mu}{}^{\nu}\}$ as
\be
\ba{rll}
H^{\mu\nu}&=&\eta^{ab}\delta_{a}^{\mu}\delta_{b}^{\nu}=\cH^{\mu\nu}\,,\\
K_{\mu\nu} &=&\delta_{\mu}^{a}\delta_{\nu}^{b}\eta_{ab}=\cH_{\mu\nu}\,,\\
Z_{\mu}{}^{\nu}&=&\delta_{\mu}^{j}\delta_{j}^{\nu}-
\delta_{\mu}^{\brj}\delta_{\brj}^{\nu}={\cH_{\mu}{}^{\nu}} \,.
\ea \label{flatspacelimit}
\ee
Parametrising  the doubled vector as 
\be
{\cV_{A}=(\varphi^{\mu},A_{\nu})}\,,
\ee 
the projectors and
the semi-covariant Yang--Mills field strength~(\ref{cF}) 
read, respectively,
\be
\ba{c}
P_{A}{}^{B}=\frac{1}{2}(\delta_{A}{}^{B}{+\cH_{A}{}^{B}})=\frac{1}{2}\!\left(\ba{cc}\delta^{\mu}{}_{\nu}+Z_{\nu}{}^{\mu}&H^{\mu\sigma}\\
K_{\rho\nu}&\delta_{\rho}{}^{\sigma}+{Z_{\rho}{}^{\sigma}}\ea\right),\\
\brP_{A}{}^{B}=\frac{1}{2}(\delta_{A}{}^{B}{-\cH_{A}{}^{B}})=\frac{1}{2}\!\left(\ba{cc}\delta^{\mu}{}_{\nu}-Z_{\nu}{}^{\mu}&-H^{\mu\sigma}\\
-K_{\rho\nu}&\delta_{\rho}{}^{\sigma}-{Z_{\rho}{}^{\sigma}}\ea\right),\\
\cF_{AB}=2\partial_{[A}\cV_{B]}-i[\cV_{A},\cV_{B}]
=\!\left(\ba{cc}
-i[\varphi^{\mu},\varphi^{\nu}]&-D_{\sigma}\varphi^{\mu}\\
~D_{\rho}\varphi^{\nu}&~f_{\rho\sigma}
\ea
\right),
\ea
\ee
where  $D_{\mu}=\partial_{\mu}-i[A_{\mu},\,\cdot\,\,]$ and
\be
f_{\mu\nu}=\partial_{\mu}A_{\nu}-\partial_{\nu}A_{\mu}-i[A_{\mu},A_{\nu}]
\label{usualf}
\ee
is the field strength of ordinary undoubled Yang--Mills theory.
From these ingredients 
 we obtain the fully  covariant field strength,
\be
\ba{c}
(P\cF\brP)_{AB}=\left(\ba{cc}
(P\cF\brP)^{\mu\nu}&(P\cF\brP)^{\mu}{}_{\sigma}\\
(P\cF\brP)_{\rho}{}^{\nu}&(P\cF\brP)_{\rho\sigma}
\ea
\right)\,,\\
\ba{rll}
(P\cF\brP)^{\mu\nu}&=&-\quarter\wtf^{\kappa\lambda}(\delta_{\kappa}{}^{\mu}+Z_{\kappa}{}^{\mu})(\delta_{\lambda}{}^{\nu}-Z_{\lambda}{}^{\nu})\,,\\
(P\cF\brP)^{\mu}{}_{\sigma}&=&\quarter\left[
H^{\mu\kappa}(\delta_{\sigma}{}^{\lambda}-Z_{\sigma}{}^{\lambda})\whf_{\kappa\lambda}-\Upsilon_{\sigma}{}^{\mu}\right]\,,\\
(P\cF\brP)_{\rho}{}^{\nu}&=&-\quarter\left[
(\delta_{\rho}{}^{\kappa}+Z_{\rho}{}^{\kappa})H^{\nu\lambda}\whf_{\kappa\lambda}-\brUpsilon_{\rho}{}^{\nu}\right]\,,\\
(P\cF\brP)_{\rho\sigma}&=&
\quarter(\delta_{\rho}{}^{\kappa}+Z_{\rho}{}^{\kappa})
(\delta_{\sigma}{}^{\lambda}-Z_{\sigma}{}^{\lambda})\whf_{\kappa\lambda}\,.
\ea
\ea \label{FYMcovariant}
\ee
Here we have introduced the shorthand notation
\be
\ba{rlll}
\wtf^{\mu\nu}&=&f^{ab}\delta_{a}{}^{\mu}\delta_{b}{}^{\nu}
+i[\varphi^{\mu},\varphi^{\nu}]-(\delta_{a}{}^{\mu}D^{a}\varphi^{\nu}+\delta_{a}{}^{\nu}D^{a}\varphi^{\mu})\,,\\
\whf_{\mu\nu}&=&f_{\mu\nu}
+i\delta_{\mu}{}^{a}\delta_{\nu}{}^{b}
[\varphi_{a},\varphi_{b}]-(\delta_{\mu}{}^{a}D_{\nu}\varphi_{a}+\delta_{\nu}{}^{a}D_{\mu}\varphi_{a})\,,\\
\Upsilon_{\mu}{}^{\nu}&=&
2\delta_{\mu}{}^{a}D^{-}_{a}\varphi^{i}\delta_{i}{}^{\nu}+4\delta_{\mu}{}^{\bri}D_{\bri}\varphi^{i}\delta_{i}{}^{\nu}\,,\\
\bar{\Upsilon}_{\mu}{}^{\nu}&=&
2\delta_{\mu}{}^{a}D^{+}_{a}\varphi^{\bri}\delta_{\bri}{}^{\nu}+4\delta_{\mu}{}^{i}D_{i}\varphi^{\bri}\delta_{\bri}{}^{\nu}\,,
\ea
\label{SHN2}
\ee
and defining $A^{\pm}_{a}= A_{a}\pm\varphi_{a}$ we further set
\be
\ba{rll}
D^{\pm}_{a}&=&\partial_{a}-i[A^{\pm}_{a},\,\cdot\,\,]\,,\\ 
f^{\pm}_{ai}&=&\partial_{a}A_{i}-\partial_{i}A^{\pm}_{a}-i[A^{\pm}_{a},A_{i}]=f_{ai}\mp D_{i}\varphi_{a}\,,\\
f^{\pm}_{a\bri}&=&\partial_{a}A_{\bri}-\brpartial_{\bri}A^{\pm}_{a}-i[A^{\pm}_{a},A_{\bri}]=f_{a\bri}\mp D_{\bri}\varphi_{a}\,.
\ea
\ee
Substituting these expressions into the Lagrangian~(\ref{YM}), after expanding as
\be
L_{\YM}
=
2\Tr\left[(P\cF\brP)^{\mu\nu}(P\cF\brP)_{\mu\nu}+(P\cF\brP)^{\mu}{}_{\nu}(P\cF\brP)_{\mu}{}^{\nu}\right]\,,
\label{usefulL}
\ee
one arrives at the undoubled Lagrangian,
\be
\! L_{\YM}^{(n,\brn)}\!=\!
\Tr\!\left[\!\ba{l}
-\quarter (f_{ab}{+i[\varphi_{a},\varphi_{b}]})(f^{ab}{+i[\varphi^{a},\varphi^{b}]})\\
-\quarter u_{ab}u^{ab}
-f^{-}_{ai}D^{-a}\varphi^{i}
+f^{+}_{a\bri}D^{+a}\varphi^{\bri}\\
-2D_{i}\varphi^{\bri}D_{\bri}\varphi^{i}
-2if_{i\bri}[\varphi^{i},\varphi^{\bri}]
\ea\!
\right]\,,
\label{sYM}
\ee
where we have defined a symmetric tensor
\be
u_{ab}=D_{a}\varphi_{b}+D_{b}\varphi_{a}\,. \label{strainflat}
\ee
Identifying}  $\varphi^{\mu}$ as the displacement  vector in elasticity theory,  $u_{ab}$ corresponds to a  strain tensor which now interacts with the undoubled Yang--Mills.  The  symmetric strain tensor    originates essentially from the projection
\be
\scalebox{0.95}{$
4({P\cF\brP})_{ab}
=f_{ab}{+i[\varphi_{a},\varphi_{b}]}{-u_{ab}}=\partial_{a}A^{-}_{b}{-\partial_{b}A^{+}_{a}}{-i[A^{+}_{a},A^{-}_{b}]}\,.$}
\ee
Since $A_{\mu}$ and $\varphi^{\mu}$ are dual to each other \textit{\`{a} la} Buscher~\cite{Buscher:1987sk,Buscher:1987qj}, so are their (Abelian) elementary quanta,   photon and phonon (\textit{c.f.~}\cite{Guddala2021}).   

By construction from   (\ref{YM}),  the Lagrangian~(\ref{sYM})  enjoys  `supertranslational'   Noether symmetries given  \textit{a priori} by $\delta\cV_{A}=\hcL_{\xi}\cV_{A}$~(\ref{gLie}) with the Killing vector~(\ref{KillingSOL}),  which   reduce in terms of  the ordinary Lie derivative  to
\be
\ba{ll}
\delta A_{\mu}=\cL_{\xi}A_{\mu}+2\partial_{[\mu}\lambda_{\rho]}\varphi^{\rho}\,,\quad&\quad
\delta\varphi^{\mu}=\cL_{\xi}\varphi^{\mu}\,.
\ea
\label{TRANSF}
\ee
In particular, under the transformations~(\ref{TRANSF}) with the Killing vector~(\ref{KillingSOL}),   $(P\cF\brP)_{AB}$ transforms covariantly as
\be
\ba{l}
\delta(P\cF\brP)^{\mu\nu}=\cL_{\xi}(P\cF\brP)^{\mu\nu}\,,\\
\delta(P\cF\brP)^{\mu}{}_{\nu}=\cL_{\xi}(P\cF\brP)^{\mu}{}_{\nu}
+2\partial_{[\nu}\lambda_{\rho]}(P\cF\brP)^{\mu\rho}\,,\\
\delta(P\cF\brP)_{\mu}{}^{\nu}=\cL_{\xi}(P\cF\brP)_{\mu}{}^{\nu}
+2\partial_{[\mu}\lambda_{\rho]}(P\cF\brP)^{\rho\nu}\,,\\
\delta(P\cF\brP)_{\mu\nu}=\cL_{\xi}(P\cF\brP)_{\mu\nu}
+2\partial_{[\mu}\lambda_{\rho]}(P\cF\brP)^{\rho}{}_{\nu} \\
\qquad\qquad\quad\;\;\; +2\partial_{[\nu}\lambda_{\rho]}(P\cF\brP)_{\mu}{}^{\rho}\,,
\ea \label{deltaPPbar}
\ee
from which  the invariance of the action,  or (\ref{usefulL}), follows straightforwardly,  
\be
\delta {L}_{\YM}=\xi^{\mu}\partial_{\mu} {L}_{\YM}=\partial_{\mu}\left(\xi^{\mu} {L}_{\YM}\right)\,.
\ee
Note that although~(\ref{deltaPPbar}) can be verified directly by brute force, it can be understood simply as a natural consequence of the $\ODD$-symmetric general covariance of the doubled Yang--Mills field strength~(\ref{FYMcovariant}), encoded  via~(\ref{gLie}).

The equations of motion ${\cD_{A}
(P\cF\brP)^{[AB]}=0}$~(\ref{EOM}) are, exhaustively, 
\be
\ba{l}
0 = D_{b}\big(F^{b}{}_{a}+i[\varphi^{b},\varphi_{a}]\big)
-D^{-}_{a}D_{i}\varphi^{i}
+D^{+}_{a}D_{\bri}\varphi^{\bri} \\ \quad\;\;\;
+i[\varphi_{b},u^{b}{}_{a}]+2i[\varphi^{i},f^{-}_{ai}]
-2i[\varphi^{\bri},f^{+}_{a\bri}]\,,\\
0=D^{-}_{a}D^{-a}\varphi^{i}+4i[\varphi^{\bri},D_{\bri}\varphi^{i}]-2i[\varphi^{i},D_{\bri}\varphi^{\bri}]\,,\\
0=D^{+}_{a}D^{+a}\varphi^{\bri}-4i[\varphi^{i},D_{i}\varphi^{\bri}]+2i[\varphi^{\bri},D_{i}\varphi^{i}]\,,\\
0=D_{b}u^{b}{}_{a}{+D^{-}_{a}}D_{i}\varphi^{i}{+D^{+}_{a}}D_{\bri}\varphi^{\bri}
{-2i\big(}[\varphi^{i},f^{-}_{ai}]+[\varphi^{\bri},f^{+}_{a\bri}]\big)\,, \\ 
0=D^{-a}f^{-}_{ai}+2D_{i}D_{\bri}\varphi^{\bri}+4i[f_{i\bri},\varphi^{\bri}]\,,\\
0=D^{+a}f^{+}_{a\bri}-2D_{\bri}D_{i}\varphi^{i}+4i[f_{i\bri},\varphi^{i}]\,.
\ea
\ee
We directly verified  the conservation ${\partial_{\mu}\vJ^{\mu}=0}$~(\ref{vJ}) by  checking, first of all, the conservation of the  doubled energy-momentum tensor itself  (up to the Bianchi identity,  the on-shell  equations~(\ref{EOM}), and  the section condition),   
\be
\ba{l}
\scalebox{0.95}{$\partial_{B}T^{B}{}_{A}=\Tr\Big[6(P\cF\brP)^{BC}D_{[A}\cF_{BC]}-4\cF_{AC}D_{B}(P\cF\brP)^{[BC]}\Big]$}\\
\qquad\scalebox{0.95}{$+4\partial_{C}\Tr\Big[\partial^{C}\cV^{B}(P\cF\brP)_{[BA]}-\cV^{C}D^{B}(P\cF\brP)_{[BA]}\Big]$}=0\,,
\ea
\label{CONSERVEDT}
\ee
and secondly,  the  vanishing of each of
\[
\big\{\,T^{ij}\,,\, T^{\bri\brj}\,,\, T^{ia}+\eta^{ab}T^{i}{}_{ b}\,,\, 
T^{\bri a}-\eta^{ab}T^{\bri}{}_{ b}\,, \,T^{i\bri}+T^{\bri i}\,\big\}
\]  
as  consequences  of the orthogonal  projections  performed in (\ref{YMT}), with  $T^{i}{}_{j}=\delta^{i}{}_{j}L_{\YM}$ and $T^{\bri}{}_{\brj}=\delta^{\bri}{}_{\brj}L_{\YM}$, \textit{c.f~}(\ref{particleT}).
Unlike the previous particle and scalar field theory cases,  both  $T^{\mu\nu}$ and $T^{\mu}{}_{\nu}$ turn out to be nontrivial in Yang--Mills theory: see {Appendix~\ref{SECA}} for the full expression, or the upcoming eq.~(\ref{TTYM}) for the simple $(0,0)$  Abelian case.

As already mentioned, it is a rather unexpected and remarkable  result of the DFT formalism that the doubled Yang--Mills theory produces a unifying description~(\ref{sYM}) of  ordinary Yang--Mills theory and a non-Abelian elasticity theory. 
We emphasize that this holds true even upon genuine  Riemannian backgrounds, \textit{i.e.~}$(n,\brn)=(0,0)$. The Abelian reduction of (\ref{sYM}) on a flat Minkowskian spacetime is simply the sum of   Maxwell and strain tensor theories describing free photons and phonons, respectively,
\be
L_{\scriptstyle{\rm{Maxwell-Strain}}}^{(0,0)}=-\quarter f_{ab}f^{ab}-\quarter u_{ab}u^{ab}\,.
\label{00MaxStr}
\ee
The corresponding energy-momentum tensors are
\be
\ba{l}
T^{ab}=2\partial_{c}\varphi^{[a}f^{b]c}+\partial_{c}(f^{ab}\varphi^{c})\,,\\
T^{a}{}_{b}=f^{ac}f_{bc}+\partial^{a}\varphi^{c}\partial_{b}\varphi_{c}-\partial_{c}\varphi^{a}\partial^{c}\varphi_{b}+\partial_{c}(\varphi^{c}u^{a}{}_{b})\\
\qquad~\quad-\quarter\delta^{a}{}_{b}\!\left(f_{cd}f^{cd}+u_{cd}u^{cd}\right)\,.
\ea
\label{TTYM}
\ee
While clearly $T^{ab}\neq T^{a}{}_{c}\eta^{cb}$, these  satisfy the off-shell relations
\be
\ba{l}
\partial_{a}T^{ab}=2\partial_{c}\big(\partial_{a}f^{a[b}\varphi^{c]}\big)\,,\\
\partial_{a}T^{a}{}_{b\!}=
f_{bd}\partial_{c}f^{cd}{-\textstyle{\frac{3}{2}}f^{cd}}\partial_{[b}f_{cd]}{+\partial_{b}\varphi^{d}}\partial_{c}u^{c}{}_{d}{+\partial_{d}}(\varphi^{d}\partial_{c}u^{c}{}_{b})\,,
\ea
\ee
and thus become conserved on-shell due to  the equations of motion, $\partial_{a}f^{ab}=0=\partial_{a}u^{ab}$.

Intriguingly, with the decomposition of the displacement vector into  temporal and spatial components,  $\varphi^{a}=(\varphi^{t},\varphi^{\aa})$ where $\aa=1,2,\cdots,D{-1}$, if we truncate the temporal component by setting $\varphi^{t}=0$, the strain tensor part in (\ref{00MaxStr}) becomes
\be
-\quarter u_{ab}u^{ab}=\half\dot{\varphi}_{\aa}\dot{\varphi}^{\aa}-\quarter u_{\aa\bb}u^{\aa\bb}\,,
\ee
where the dots denote time derivatives~$\frac{\rd~}{\rd t}$. This is precisely the elasticity theory   considered in \cite{Pretko:2017kvd},  shown therein to be dual to a  $\mathbf{U}(1)$ symmetric tensor gauge theory. The topological defects of the former  map onto charges of the latter, with disclinations and dislocations corresponding to fractons and dipoles, respectively.  Our result is in some sense (orthogonally) complementary to \cite{Pretko:2017kvd}, as  we are principally focusing on the fracton physics that arises geometrically  from non-Riemannian backgrounds. Nevertheless, we remark that both non-Riemannian spaces and topological defects are, after all, singular configurations from a conventional perspective. \\

\section{Doubled Maxwell Coupled  to Charged Particles\label{SECMP}}
We now consider the doubled Maxwell theory~(\ref{YM})  minimally coupled to    particles~(\ref{particle}) with charge $q$,
\be
\sum_{\alpha}
\dis{\int}\rd\tau~\half e^{-1}\gD_{\tau}x_{\alpha}^{A}\gD_{\tau}x_{\alpha}^{B}\cH_{AB}-\half em^{2}-q\gD_{\tau}x_{\alpha}^{A}\cV_{A}\,,
\label{minimalC}
\ee
where $\alpha$ denotes a (negligible) particle index.   On the constant $(n,\brn)$ background~(\ref{flatH}),  the single-particle Lagrangian becomes
\be
\ba{lll}
\cL_{q}&=&{\frac{1}{2}}e^{-1}\gD_{\tau}x^{A}\gD_{\tau}x^{B}\cH_{AB}-\half em^{2}-q\gD_{\tau}x^{A}\cV_{A}\\
&=&{\frac{1}{2}}e^{-1}\dot{x}^{a}\dot{x}_{a}-\half e\big(m^{2}+q^{2}\varphi^{a}\varphi_{a}\big)-q\dot{x}^{\mu}A_{\mu}\\
{}&{}&\!\!\!\!\!
+{\frac{1}{2}}e^{-1}
\big(\dot{\tx}_{a}-\mathbf{a}_{a}
-eq\varphi_{a}\big)\big(\dot{\tx}_{b}-\mathbf{a}_{b}
-eq\varphi_{b}\big)\eta^{ab}\\
{}&{}&\!\!\!\!\!
+\big(e^{-1}\dot{y}^{i}-q\varphi^{i}\big)\big(
\dot{\ty}_{i}-\mathbf{a}_{i}\big)
-\big(e^{-1}\dot{\bry}^{\bri}+q\varphi^{\bri}\big)\big(
\dot{\tilde{\bry}}_{\bri}-\mathbf{a}_{\bri}\big)\,.
\ea
\label{intermediate}
\ee
The corresponding action generalises (\ref{Fracton}) to
\be
S_{q}=\dis{\int}\rd\tau\left[\ba{l}
\half e^{-1}\dot{x}^{a}\dot{x}_{a}  -\half e(m^{2}+q^{2}\varphi^{a}\varphi_{a}) -q\dot{x}^{\mu}A_{\mu}\\
+\Lambda_{i}(\dot{y}^{i}-eq\varphi^{i})+\brLambda_{\bri}(\dot{\bry}^{\bri}+eq\varphi^{\bri})\ea\right]\,,
\label{SFracton2}
\ee
where we have integrated out the auxiliary variables $\mathbf{a}_{a}$ thereby setting the on-shell value $\gD_{\tau}\tx_{a}=eq\varphi_{a}$, 
and we have identified ${\Lambda_{i}=e^{-1}\gD_{\tau}\ty_{i}}$ and ${\brLambda_{\bri}=-e^{-1}\gD_{\tau}\tilde{\bar{y}}_{\bri}}$.
This action then couples the particle to a strain-Maxwell theory, \textit{i.e.~}the Abelian reduction  of (\ref{sYM}),
\be
L_{{\scriptscriptstyle{0}}}=-\quarter f_{ab}f^{ab}
-\quarter u_{ab}u^{ab}
-f^{-}_{ai}\partial^{a}\varphi^{i}
+f^{+}_{a\bri}\partial^{a}\varphi^{\bri}
-2\partial_{i}\varphi^{\bri}\partial_{\bri}\varphi^{i}\,.
\label{strainMaxwell}
\ee 
The Hamiltonian action for (\ref{SFracton2}) is
\be
\ba{c}
S_{H}=\dis{\int}\rd\tau~p_{\mu}\dot{x}^{\mu}-eH\,,\\
\ba{lll}
H&=&\half(p_{a}+qA_{a})(p^{a}+qA^{a})+\half(m^{2}+q^{2}\varphi_{a}\varphi^{a})\\{}&{}&+q(p_{i}+qA_{i})\varphi^{i}-q(\brp_{\bri}+qA_{\bri})\varphi^{\bri}\,.
\ea
\ea
\label{SH}
\ee
Integrating out the $p_{a}$'s from (\ref{SH}), one recovers (\ref{SFracton2}) with the identification $p_{i}=\Lambda_{i}-qA_{i}$, $\brp_{\bri}=\brLambda_{\bri}-qA_{\bri}$.

Clearly,  from  (\ref{SFracton2}) or (\ref{SH}), the immobility  is  lifted by the  displacement  vector to a `saturation velocity',
\be
\ba{ll}
\dot{y}^{i}=eq\varphi^{i}\,,\qquad&\qquad
\dot{\bry}^{\bri}=-eq\varphi^{\bri}\,.
\ea
\label{velocitysaturation}
\ee 
Further, the Hamiltonian constraint $H=0$~(\ref{SH}) leads to a modified dispersion relation,
\be
k_{a}k^{a}+m^{2}+q^{2}\varphi_{a}\varphi^{a}+2qk_{i}\varphi^{i}-2q\bar{k}_{\bri}\varphi^{\bri}=0\,,
\ee
in which one may  identify an effective mass  (\textit{c.f.~}{\cite{Bulmash:2018lid, Ma:2018nhd}} for an interpretation in terms of a Higgs mechanism),
\be
m_{{\rm{eff.}}}^{2}=m^{2}+q^{2}\varphi^{a}\varphi_{a}\,.
\label{effMASS}
\ee  
We stress that while  (\ref{velocitysaturation}) is a consequence of the  non-Riemannian geometry,  (\ref{effMASS}) holds on the Riemannian subspace.

The equations of motion of the photon $A_{\mu}$  are the following generalised Maxwell equations,
\be
\ba{l}
\partial_{b}f^{ba}-\partial^{a}\partial_{i}\varphi^{i}+\partial^{a}\brpartial_{\bri}\varphi^{\bri}=J^{a}\,,\\
\partial_{b}\partial^{b}\varphi^{i}=J^{i}\,,\quad\,
-\partial_{b}\partial^{b}\varphi^{\bri}=J^{\bri}\,.
\ea
\label{MaxwellEQ}
\ee
Here $J^{\mu}$ is the usual  electric current,
\be
J^{\mu}(x)=\dis{\sum_{\alpha}\int\rmd\tau}~q\dot{x}^{\mu}_{\alpha}(\tau)\delta^{D\!}\big(x-x_{\alpha}(\tau)\big)\,,
\label{Jmu}
\ee
which is    identically conserved in the manner of  (\ref{identicalCONSERVATION}),  ensuring  the consistency of (\ref{MaxwellEQ})  \textit{\`{a} la} Maxwell in 1865.   

Meanwhile, the phonon $\varphi^{\mu}$ has equations of motion
\be
\ba{r}
\partial_{b}u^{b}{}_{a}+\partial_{a}\partial_{i}\varphi^{i}+\partial_{a}\brpartial_{\bri}\varphi^{\bri}=\tJ_{a}\,,\\
\partial_{i}\partial_{b}\varphi^{b}+2\partial_{i}\brpartial_{\bri}\varphi^{\bri}+\partial_{b}f^{b}{}_{i}=\tJ_{i}\,,\\
\brpartial_{\bri}\partial_{b}\varphi^{b}+2\brpartial_{\bri}\partial_{i}\varphi^{i}-\partial_{b}f^{b}{}_{\bri}=\tJ_{\bri}\,,
\ea
\label{strainEQ}
\ee
for which  we introduce  a  (by no means conserved)  dual `pseudo-current',
\be
\tJ_{\mu}(x)=\dis{\sum_{\alpha}\int\rmd\tau}~q\gD_{\tau}\tx_{\alpha\mu}(\tau)\delta^{D\!}\big(x-x_{\alpha}(\tau)\big)\,.
\label{tJpseudo}
\ee
On shell, in terms of  $\varphi_{a}$ and  the conjugate momenta $p_{i},\brp_{\bri}$ of $y^{i},\bry^{\bri}$, we can 
express the dual pseudo-current without  explicitly  invoking       the tilde coordinates,
\be
\ba{l}
\tJ_{a}(x)=\dis{\sum_{\alpha}\int\rmd\tau}~eq^{2}\varphi_{a\,}\delta^{D\!}\big(x-x_{\alpha}(\tau)\big)\,,\\
\tJ_{i}(x)=\dis{\sum_{\alpha}\int\rmd\tau}~eq(p_{\alpha i}+qA_{i})\,\delta^{D\!}\big(x-x_{\alpha}(\tau)\big)\,,\\
\tJ_{\bri}(x)=\dis{\sum_{\alpha}\int\rmd\tau}~(-eq)(\brp_{\alpha \bri}+qA_{\bri})\,\delta^{D\!}\big(x-x_{\alpha}(\tau)\big)\,.
\ea
\label{pseudoJ}
\ee
Evidently, (\ref{MaxwellEQ}) and (\ref{strainEQ})  generalise the usual  Maxwell equations. In particular, in the absence of  sources, they describe  electromagnetic-strain \textit{waves} that propagate exclusively through  the Riemannian subspace, and  thus are  fractonic,
\be
\partial_{c}\partial^{c}f_{\mu\nu\!}=
\partial_{c}\partial^{c}\varphi^{i\!}=
\partial_{c}\partial^{c}\varphi^{\bri\!}=
\partial_{c}\partial^{c}\partial_{a}\varphi^{a\!}=
\partial_{c}\partial^{c}\partial_{[a}\varphi_{b]\!}=0\,.
\label{fractonic}
\ee
The  first term vanishes on-shell via the Bianchi identity as ${\partial_{c}\partial^{c}f_{\mu\nu}=-2\partial_{[\mu}\partial^{c}f_{\nu]c}=0}$, while the final equality,  which pertains to  the `rotation tensor'  $\partial_{[a}\varphi_{b]}$, follows from    the  more general relation 
${\partial_{c}\partial^{c}\varphi_{a}+\partial_{a}\partial_{\lambda}\varphi^{\lambda}=0}$.

The consistent matching between the particle action, (\ref{particle}) or (\ref{Fracton}),  and the scalar field theory~(\ref{LPhi})   generalises to the interacting theory of a charged particle~(\ref{intermediate}) and a charged complex scalar field. The Lagrangian of the complex scalar field in the fundamental representation  is,   with $\fD_{A}\Phi=(\partial_{A}+iq\cV_{A})\Phi$ and  
$\fD_{\mu}\Phi=(\partial_{\mu}+iqA_{\mu})\Phi$, 
\be
\ba{lll}
L_{\cV,\Phi}\!\!&=\!\!&-\half\cH^{AB}\fD_{A}\Phi \fD_{B}\Phi^{\dagger}-\half m^{2}\Phi\Phi^{\dagger}\\
{}\!\!&=\!\!&
-\half \fD_{a}\Phi\fD^{a}\Phi^{\dagger}-\half(m^{2}+q^{2}\varphi_{a}\varphi^{a})\Phi\Phi^{\dagger}\\
{}\!\!&\!\!&
-i\half q\varphi^{i}\big(\Phi\fD_{i}\Phi^{\dagger}{-\Phi^{\dagger}\fD_{i}\Phi}\big)
\\
{}\!\!&\!\!&
+i\half q\varphi^{\bri}\big(\Phi\fD_{\bri}\Phi^{\dagger}{-\Phi^{\dagger}\fD_{\bri}\Phi}\big)\,.
\ea
\label{LPhibarPhi}
\ee
Remarkably, the  resulting  equation of motion  agrees with the Hamiltonian constraint of the charged particle~(\ref{SH}), including the mass enhancement,
\be
\Big[
{-\fD_{a}\fD^{a}}+m^{2}{+q^{2}\varphi_{a}\varphi^{a}}-iq\{\fD_{i},\varphi^{i}\}+iq\{\fD_{\bri},\varphi^{\bri}\}\Big]\Phi=0\,,
\ee
while it further produces a symmetric `prescription' for the  Hamiltonian constraint at the quantum level,
\[
\ba{ll}
\{p_{i},\varphi^{i}\}=p_{i}\varphi^{i}+\varphi^{i}p_{i}\,,\quad&\quad
\{\brp_{\bri},\varphi^{\bri}\}=\brp_{\bri}\varphi^{\bri}+\varphi^{\bri}\brp_{\bri}\,,
\ea
\]
after identifying $-i\fD_{\mu}=-i\partial_{\mu}+qA_{\mu}$, with $p_{\mu}+qA_{\mu}$.\\

\section{On Curved Non-Riemmannian Backgrounds\label{SECCurved}}
Owing to the geometric $\ODD$ formalism applicable to non-Riemannian geometries that has been developed in the literature~\cite{Morand:2017fnv,Cho:2019ofr},  all results in the previous sections can be readily generalised to curved backgrounds.  In the context of DFT this also includes the possibility of a non-trivial $B$-field and dilaton.  Here we present such curved results in full generality: in doing so it is necessary to define many covariant quantities while carefully distinguishing upper and lower $D$-dimensional curved indices, $\mu,\nu$.  Though this may at first glance seem over-elaborate, we remind and warn the reader that raising and lowering indices is generically not possible in the absence of an invertible metric.

 Ref.\cite{Slagle:2018kqf} and two recent works \cite{Bidussi:2021nmp,Jain:2021ibh} already   considered  fracton physics on curved backgrounds using sub-Riemannian or Carrollian/Aristotelian geometries, which in the present framework would correspond to the $(1,0)$ or  $(D{-1},0)$ non-Riemannian geometries.   Such scenarios, not being  organised under   the generalised metric,  break    $\ODD$ symmetry.     The consequences of this remain to be seen.  Here  we merely present our $\ODD$  symmetric, curved spacetime  extension of the  previous  particle, scalar field, and Yang--Mills theories.

 On general curved backgrounds, it is convenient to factorise out the $B$-field contribution via an $\ODD$ transformation,
\be
\ba{ll}
\cB_{A}{}^{B}
=\left(\ba{cc}\delta^{\mu}{}_{\sigma}&0\\B_{\rho\sigma}&~
\delta_{\rho}{}^{\tau}\ea\right),
\qquad&\qquad
\cB_{A}{}^{C}\cB_{B}{}^{D}\cJ_{CD}=
\cJ_{AB}\,.
\ea
\label{cB}
\ee
The $\ODD$-covariant generalised metric 
on a general curved background then takes the form \cite{Morand:2017fnv}
\be
\ba{ll}
\cH_{AB}=\cB_{A}{}^{C}\cB_{B}{}^{D}\mcH_{CD}
\,,\qquad&\qquad
\mcH_{AB}=\left(\ba{cc}
H^{\mu\nu}&
Z_{\lambda}{}^{\mu}\\
Z_{\kappa}{}^{\nu}&
K_{\kappa\lambda}
\ea\right)\,,
\ea
\label{nnbar}
\ee
where, with  ${1\leq i \leq n}$,  ${1\leq\bri\leq\brn}$ as before,
\be
Z_{\mu}{}^{\nu} = X^{i}_{\mu}Y_{i}^{\nu}
-\brX^{\bri}_{\mu}\brY_{\bri}^{\nu} \, .
\ee
The vectors $X^{i}_{\mu}, \brX^{\bri}_{\mu}$ and $Y_{i}^{\mu}, \brY_{i}^{\mu}$ belong to the kernels of $H^{\mu\nu}$ and $K_{\mu\nu}$, respectively,
\be
H^{\mu\nu}X^{i}_{\nu} = 0 = H^{\mu\nu}\brX^{\bri}_{\nu} \,, \qquad K_{\mu\nu}Y_{i}^{\nu} = 0 = K_{\mu\nu}\brY_{\bri}^{\nu} \,,
\label{zero}
\ee
and correspond to the $n+\brn$ non-Riemannian directions.  These objects satisfy a completeness relation,
\be
H^{\mu\rho}K_{\rho\nu} + Y_{i}^{\mu}X^{i}_{\nu} + \brY_{\bri}^{\mu}\brX^{\bri}_{\nu} = \delta^{\mu}{}_{\nu} \,.
\label{completeness}
\ee
From (\ref{zero}), (\ref{completeness}), and the linear independence of the null eigenvectors,   it follows that
\be
\ba{c}
X^{i}_{\mu}Y^{\mu}_{j}=\delta^{i}_{j}\,,\quad\quad
\brX^{\bri}_{\mu}\brY^{\mu}_{\brj}=\delta^{\bri}_{\brj}\,,\quad\quad
X^{i}_{\mu}\brY^{\mu}_{\bri}=0=\brX^{\brj}_{\mu}Y^{\mu}_{j}\,,\\
(HKH)^{\mu\nu}=H^{\mu\nu}\,,\quad\quad
(KHK)_{\mu\nu}=K_{\mu\nu}\,.
\ea
\ee
Further from (\ref{cB}) and (\ref{nnbar}),   a similar factorisation of the projectors holds,
\be
P_{AB} = \cB_{A}{}^{C}\cB_{B}{}^{D}\mP_{CD}\,,\qquad \brP_{AB} = \cB_{A}{}^{C}\cB_{B}{}^{D}\mbrP_{CD}\,,\label{mPmbrP}
\ee
where $\mP_{AB} = \half(\cJ + \mcH)_{AB}$ and $\mbrP_{AB} = \half(\cJ - \mcH)_{AB}$.

Remarkably, the doubled metric $\cH_{AB}$ is invariant under   $\GL(n)\times\GL(\brn)$  local rotations,  which  act  on    the unbarred $i,j,\cdots$ and barred $\bri,\brj,\cdots$ indices, and further under the  `Milne-shift' symmetry---generalising  the   `Galilean  boost' in the Newtonian  gravity literature~\cite{Milne:1934,Duval:1993pe}---which  acts with arbitrary local parameters, $V_{\mu i}$ and  $\brV_{\mu\bri}$,  as~\cite{Morand:2017fnv}
\be
\ba{c}
\deltaM H^{\mu\nu}=0\,,\qquad\deltaM X^{i}_{\mu}=0\,,\qquad\deltaM \brX_{\mu}^{\bri}=0\,,\\
\deltaM Y^{\mu}_{i} =H^{\mu\nu}V_{\nu i}\,,\qquad
\deltaM\brY_{\bri}^{\mu}=H^{\mu\nu}\brV_{\nu\bri}\,,\\ 
\deltaM K_{\mu\nu}= -2X^{i}_{(\mu}K_{\nu)\rho}H^{\rho\sigma}V_{\sigma i}-2\brX^{\bri}_{(\mu}K_{\nu)\rho}H^{\rho\sigma}\brV_{\sigma\bri}
\,,\\
\deltaM B_{\mu\nu}=
-2X^{i}_{[\mu}V_{\nu]i}+2\brX^{\bri}_{[\mu}\brV_{\nu]\bri}\\
\qquad\qquad\qquad\qquad+2X^{i}_{[\mu}\brX^{\bri}_{\nu]}\left(Y_{i}^{\rho}\brV_{\rho\bri}
+\brY_{\bri}^{\rho}V_{\rho i}
\right).
\ea
\label{Milne1}
\ee
In fact, both local symmetries are part of the local Lorentz symmetries in DFT and should not be broken.

First let us briefly comment on the scalar field case.  Upon the generic $(n,\brn)$ curved background  above, with the choice of section~${\tpartial^{\mu}=0}$, the scalar field  kinetic term reduces to    
\be
-\half e^{-2d}\cH^{AB}\partial_{A}\Phi\partial_{B}\Phi=-\half e^{-2d}H^{\mu\nu}\partial_{\mu}\Phi\partial_{\nu}\Phi\,,
\ee
which obviously generalises  (\ref{LPhi}) into a covariant form.

Now we turn to the doubled Yang--Mills theory on curved backgrounds.  We  first  factorise the doubled gauge potential, similarly to  (\ref{nnbar}), as
\be
\ba{ll}
\cV_{A}=\cB_{A}{}^{B}\mcV_{B}=\left(\ba{c}\varphi^{\mu}\\
A_{\nu}+B_{\nu\rho}\varphi^{\rho}\ea\right)\,,\quad&\quad\mcV_{A}=
\left(\ba{c}\varphi^{\mu}\\
A_{\nu}\ea\right)\,.
\ea
\ee
Like the doubled metric, the doubled vector potential $\cV_{A}$ should be invariant under the DFT local Lorentz symmetries,  and thus the Milne-shift transformations of the component fields  are
\be
\ba{ll}
\deltaM\varphi^{\mu}=0\,,\qquad&\qquad\deltaM A_{\mu}=-(\deltaM B_{\mu\nu})\varphi^{\nu}\,.
\ea
\label{Milne2}
\ee

The semi-covariant Yang--Mills field strength is
\be
\ba{rl}
\cF_{AB}=&\!\na_{A}\cV_{B}-\na_{B}\cV_{A}-i\left[\cV_{A},\cV_{B}\right]\\=&\!2\partial_{[A}\cV_{B]}+2\Gamma_{[AB]}{}^{C}\cV_{C}
-i\left[\cV_{A},\cV_{B}\right]\,, \label{cFcurved}
\ea
\ee
where $\Gamma_{CAB}$ are the the DFT Christoffel symbols  \cite{Jeon:2011cn}.  From the torsionless property $\Gamma_{[ABC]} = 0$, it follows that $\Gamma_{[AB]C} = -\half\Gamma_{CAB}$.  Since the fully covariant field strength is $(P\cF\brP)_{AB}$, we only need the projection
\be
P_{M}{}^{A}\brP_{N}{}^{B}\Gamma_{CAB}=\left(P\partial_{C}P\brP\right)_{MN} = \half\left(P\partial_{C}\cH\brP\right)_{MN}\,.
\label{GammaoPbrP}
\ee

Using~(\ref{cB}),~(\ref{nnbar}),~(\ref{mPmbrP}), and (\ref{GammaoPbrP})  in (\ref{cFcurved}),   the fully covariant field strength is given by
\be
\ba{l}
(\cB^{-1}P)_{A}{}^{C}(\cB^{-1}\brP)_{B}{}^{D}\cF_{CD}\\=\mP_{A}{}^{C}\mbrP_{B}{}^{D}\Big(
2\partial_{[C}\mcV_{D]}-i[\mcV_{C},\mcV_{D}]-\half\mcV^{E}\partial_{E}\mcH_{CD}\\ \qquad\qquad\quad\;\; +3\mcV^{E}\partial_{[E}\cB_{CD]}\Big)\,. \label{FYMcurvariant}
\ea
\ee
This is the curved generalisation of~(\ref{FYMcovariant}).  After a lengthy calculation we may  obtain the explicit components,
\be
\ba{l}
(\cB^{-1}P\cF(\cB^{-1}\brP)^t)^{\mu\nu}=-\quarter\wtf^{\kappa\lambda}(\delta_{\kappa}{}^{\mu}+Z_{\kappa}{}^{\mu})(\delta_{\lambda}{}^{\nu}-Z_{\lambda}{}^{\nu})\,,\\
(\cB^{-1}P\cF(\cB^{-1}\brP)^t)^{\mu}{}_{\sigma}=\quarter\left[
H^{\mu\kappa}(\delta_{\sigma}{}^{\lambda}-Z_{\sigma}{}^{\lambda})\whf_{\kappa\lambda}-\Upsilon_{\sigma}{}^{\mu}\right]\,,\\
(\cB^{-1}P\cF(\cB^{-1}\brP)^t)_{\rho}{}^{\nu}=-\quarter\left[
(\delta_{\rho}{}^{\kappa}+Z_{\rho}{}^{\kappa})H^{\nu\lambda}\whf_{\kappa\lambda}-\brUpsilon_{\rho}{}^{\nu}\right]\,,\\
(\cB^{-1}P\cF(\cB^{-1}\brP)^t)_{\rho\sigma}=
\quarter(\delta_{\rho}{}^{\kappa}+Z_{\rho}{}^{\kappa})
(\delta_{\sigma}{}^{\lambda}-Z_{\sigma}{}^{\lambda})\whf_{\kappa\lambda}\,, \label{FYMcurvedcomponents}
\ea
\ee
where the constituent tensors are now defined as
\be
\ba{l}
\wtf^{\mu\nu}=H^{\mu\rho}H^{\nu\sigma}\big(f_{\rho\sigma}+H_{\rho\sigma\tau}\varphi^{\tau}\big)
+i[\varphi^{\mu},\varphi^{\nu}]-\wtu^{\mu\nu}\,,\\
\whf_{\mu\nu}=f_{\mu\nu}
+iK_{\mu\rho}K_{\nu\sigma}[\varphi^{\rho},\varphi^{\sigma}]+\varphi^{\rho}H_{\rho\mu\nu}-u_{\mu\nu}\,,\\
\Upsilon_{\mu}{}^{\nu}=
2K_{\mu\rho}\big(\cD^{\rho}\varphi^{i}+i[\varphi^{\rho},\varphi^{\sigma}]X^{i}_{\sigma}\big)Y^{\nu}_{i}
+4\brX_{\mu\,}^{\bri}\brcD_{\bri}\varphi^{\rho} X_{\rho}^{i\,}Y^{\nu}_{i}\,,\\
\bar{\Upsilon}_{\mu}{}^{\nu}=
2K_{\mu\rho}\big(\brcD^{\rho}\varphi^{\bri}-i[\varphi^{\rho},\varphi^{\sigma}]\brX^{\bri}_{\sigma}\big)\brY^{\nu}_{\bri} +4X_{\mu\,}^{i}\cD_{i}\varphi^{\rho}\brX_{\rho}^{\bri\,}\brY^{\nu}_{\bri}\,.
\ea
\label{SHN}
\ee
Further explanation is in order. Every term in (\ref{SHN}) is covariant under  (ordinary undoubled) diffeomorphisms and Yang--Mills gauge symmetry, and also invariant under $\GL(n)\times\GL(\brn)$ local rotations, but not under the Milne shift: only the whole set of components of $(P\cF\brP)_{AB}$ is so. Specifically, $f_{\mu\nu}$ is the usual Yang--Mills field strength~(\ref{usualf}),
while $\wtu^{\mu\nu}=\wtu^{\nu\mu}$, $u_{\mu\nu}=u_{\nu\mu}$ are the  \textit{strain tensors} for the vector field~$\varphi^{\lambda}$, carrying  upper or lower    symmetric indices, which can be expressed as symmetrisations of appropriate covariant derivatives,
\be
\ba{ll}
\wtu^{\mu\nu}=\fD^{\mu}\varphi^{\nu}+\fD^{\nu}\varphi^{\mu}\,,\qquad&\qquad
u_{\mu\nu}=\cD_{\mu}\varphi_{\nu}+\cD_{\nu}\varphi_{\mu}\,,
\ea \label{strain}
\ee
where, with $\Omega^{\mu\nu}{}_{\rho}$ to be explained later~(\ref{Omega}),
\be
\ba{rll}
\fD^{\mu}\varphi^{\nu}&=&H^{\mu\rho}\left(\partial_{\rho}\varphi^{\nu}
-i[A_{\rho},\varphi^{\nu}]\right)+\Omega^{\mu\nu}{}_{\rho}\varphi^{\rho}\,,\\
\cD_{\mu}\varphi_{\nu}&=&\left(\partial_{\mu}\varphi^{\rho}-i[A_{\mu},\varphi^{\rho}]\right)K_{\rho\nu}\\ & &+\half(\partial_{\rho}K_{\mu\nu}
+\partial_{\mu}K_{\nu\rho}-\partial_{\nu}K_{\mu\rho})\varphi^{\rho}\,.
\ea
\label{covD3}
\ee
Further, we  have  defined various covariant derivatives which are nontrivial only in genuine non-Riemannian cases, \textit{i.e.~}$(n,\brn)\neq(0,0)$:
\be
\ba{rll}
\cD^{\mu}\varphi^{i}&=&H^{\mu\rho}\Big(\left(\partial_{\rho}\varphi^{\sigma}-i[A_{\rho},\varphi^{\sigma}]\right)X^{i}_{\sigma}+\varphi^{\sigma}\partial_{\sigma}X^{i}_{\rho}\Big)\,,\\
\brcD^{\mu}\varphi^{\bri}&=&H^{\mu\rho}\Big(\left(\partial_{\rho}\varphi^{\sigma}
-i[A_{\rho},\varphi^{\sigma}]\right)\brX^{\bri}_{\sigma}+\varphi^{\sigma}\partial_{\sigma}\brX^{\bri}_{\rho}\Big)\,,\\
\cD_{i}\varphi^{\mu}\brX^{\bri}_{\mu}&=&\Big(Y_{i}^{\rho}\left(\partial_{\rho}\varphi^{\mu}
-i[A_{\rho},\varphi^{\mu}]\right)-\varphi^{\rho}\partial_{\rho}Y_{i}^{\mu}\Big)\brX^{\bri}_{\mu}\,,\\
\brcD_{\bri}\varphi^{\mu}X^{i}_{\mu}&=&\Big(\brY_{\bri}^{\rho}\left(\partial_{\rho}\varphi^{\mu}
-i[A_{\rho},\varphi^{\mu}]\right)-\varphi^{\rho}\partial_{\rho}\brY_{\bri}^{\mu}\Big)X_{\mu}^{i}\,.
\ea
\label{covD1}
\ee
As alternatives to the latter two expressions in~(\ref{covD1}), we can also write
\be
\ba{l}
\cD_{i}\varphi^{\lambda} K_{\lambda\mu}=\Big(Y_{i}^{\rho}\left(\partial_{\rho}\varphi^{\lambda}-i[A_{\rho},\varphi^{\lambda}]\right)
-\varphi^{\rho}\partial_{\rho}Y_{i}^{\lambda}\Big)K_{\lambda\mu}\,,\\
\brcD_{\bri}\varphi^{\lambda} K_{\lambda\mu}=\Big(\brY_{\bri}^{\rho}\left(\partial_{\rho}\varphi^{\lambda}-i[A_{\rho},\varphi^{\lambda}]\right)
-\varphi^{\sigma}\partial_{\sigma}\brY_{\bri}^{\lambda}\Big)K_{\lambda\mu}\,.
\ea
\label{covD2}
\ee

All expressions for covariant derivatives given in   (\ref{covD3}), (\ref{covD1}), and (\ref{covD2}) are symmetric under diffeomorphisms,  local $\GL(n)\times\GL(\brn)$ rotations, and Yang--Mills gauge transformations.  However, without the contractions as in (\ref{covD1}) and (\ref{covD2}), the bare derivatives
\be
\ba{rll}
\cD_{i}\varphi^{\mu}&=&Y_{i}^{\rho}\left(\partial_{\rho}\varphi^{\mu}-i[A_{\rho},\varphi^{\mu}]\right)
-\varphi^{\rho}\partial_{\rho}Y_{i}^{\mu} \,,\\ \brcD_{\bri}\varphi^{\mu}&=&\brY_{\bri}^{\rho}\left(\partial_{\rho}\varphi^{\mu}-i[A_{\rho},\varphi^{\mu}]\right)
-\varphi^{\rho}\partial_{\rho}\brY_{\bri}^{\mu}\,,
\ea \label{covDbare}
\ee
are anomalous under local $\GL(n)\times\GL(\brn)$ rotations due to the final terms containing derivatives of $Y_{i}^{\mu}$ and $\brY_{\bri}^{\mu}$.  While most of these covariant derivatives are novel, the upper-indexed covariant derivative,  
\be
\fD^{\mu}=H^{\mu\rho}\partial_{\rho}-iH^{\mu\rho}[A_{\rho},~~]+\Omega^{\mu}\,,
\ee
was proposed in \cite{Cho:2019ofr}.  It can act on an arbitrary (undoubled) tensor, as it is equipped with a generalised Christoffel connection,
\be
\ba{lll}
\Omega^{\mu\nu}{}_{\lambda}&=&-\half\partial_{\lambda}H^{\mu\nu}
-H^{\rho[\mu}\partial_{\rho}Y_{i}^{\nu]}X^{i}_{\lambda} -H^{\rho[\mu}\partial_{\rho}\brY_{\bri}^{\nu]}\brX^{\bri}_{\lambda}\\ {}&{}&- H^{\rho[\mu}\partial_{\rho}H^{\nu]\sigma}K_{\sigma\lambda}
+\Big(2
H^{\rho[\mu}Y_{i}^{\nu]}\partial_{[\tau}X^{i}_{\rho]} \\ {}&{}&
-2H^{\rho[\mu}\brY_{\bri}^{\nu]}\partial_{[\tau}\brX^{\bri}_{\rho]}
\Big)\!\left(Y_{j}^{\tau}X_{\lambda}^{j}-\brY_{\brj}^{\tau}\brX_{\lambda}^{\brj}\right).
\ea
\label{Omega}
\ee
Note that according to~(\ref{strain}), only the symmetric part, $\Omega^{(\mu\nu)}{}_{\lambda} = -\half\partial_{\lambda}H^{\mu\nu}$, contributes to the strain tensor $\wtu^{\mu\nu}$.

The Lagrangian for the doubled Yang--Mills theory on general curved and non-Riemannian backgrounds,
\be
L_{\YM}
=
2\Tr\left[(P\cF\brP)^{\mu\nu}(P\cF\brP)_{\mu\nu}+(P\cF\brP)^{\mu}{}_{\nu}(P\cF\brP)_{\mu}{}^{\nu}\right]\,,
\label{usefulLagain}
\ee
can be obtained analogously to the flat case from the components in~(\ref{FYMcurvedcomponents}).  The full result is\\

\onecolumngrid
\noindent--------------------------------------------------------------------------
\vspace{3pt}
\be
L_{\YM}\!=\Tr\!\left[\!\ba{l}
-\quarter H^{\mu\rho}H^{\nu\sigma}\big(f_{\mu\nu}{+\varphi^{\kappa}}H_{\kappa\mu\nu}\big)\big(f_{\rho\sigma}{+\varphi^{\lambda}}H_{\lambda\rho\sigma}\big)-\quarter H^{\mu\rho}H^{\nu\sigma}u_{\mu\nu}u_{\rho\sigma}
+\quarter K_{\mu\rho}K_{\nu\sigma}[\varphi^{\mu},\varphi^{\nu}][\varphi^{\rho},\varphi^{\sigma}]\\
-\Big\{\cD^{\mu}\varphi^{i}Y_{i}^{\nu}-
\brcD^{\mu}\varphi^{\bri}\brY_{\bri}^{\nu}+i\half[\varphi^{\rho},\varphi^{\sigma}] (1{+Z})_{\rho}{}^{\mu}(1{-Z})_{\sigma}{}^{\nu}\Big\}\big(f_{\mu\nu}{+\varphi^{\kappa}}H_{\kappa\mu\nu}\big)\\
-K_{\mu\nu}\cD_{i}\varphi^{\mu}\big(\cD^{\nu}\varphi^{i}
+i[\varphi^{\nu},\varphi^{\rho}]X^{i}_{\rho}\big)
-K_{\mu\nu}\brcD_{\bri}\varphi^{\mu}\big(\brcD^{\nu}\varphi^{\bri}
-i[\varphi^{\nu},\varphi^{\rho}]\brX^{\bri}_{\rho}\big)
-2\cD_{i}\varphi^{\mu}\brX^{\bri}_{\mu}\brcD_{\bri}\varphi^{\nu}X^{i}_{\nu}
\ea\!
\right]\,.
\label{Fsquare}
\ee
\vspace{3pt}
\qquad\qquad\qquad\qquad\qquad\qquad
\qquad\qquad\qquad\qquad\qquad\qquad\qquad
--------------------------------------------------------------------------
\vspace{3pt}
\twocolumngrid
\noindent By construction, this action is invariant under  diffeomorphisms, $B$-field/Yang--Mills gauge symmetries, $\GL(n)\times\GL(\brn)$ local rotations, and Milne shifts~(\ref{Milne1}), (\ref{Milne2}). This follows naturally in the doubled formalism, but appears nontrivial from the undoubled perspective.\\

 In deriving the action, it is worthwhile to note the following relations   among the derivatives $\cD^{\mu}\varphi^{i}$,  $\brcD^{\mu}\varphi^{\bri}$ and $\fD^{\mu}\varphi^{\rho}$, 
\be
\ba{l}
(\fD^{\mu}\varphi^{\rho})X^{i}_{\rho}Y^{\nu}_{i} -
(\cD^{\mu}\varphi^{i})Y_{i}^{\nu} \\ 
=H^{\mu\rho}\partial_{[\rho}X^{i}_{\sigma]}\big(H^{\sigma\kappa}K_{\kappa\lambda}
+2\brY^{\sigma}_{\bri}\brX^{\bri}_{\lambda}\big)\varphi^{\lambda}Y_{i}^{\nu}\,,\\
(\fD^{\mu}\varphi^{\rho})\brX^{\bri}_{\rho}\brY^{\nu}_{\bri} -
(\brcD^{\mu}\varphi^{\bri})\brY_{\bri}^{\nu} \\ 
=H^{\mu\rho}\partial_{[\rho}\brX^{\bri}_{\sigma]}\big(H^{\sigma\kappa}K_{\kappa\lambda}
+2Y^{\sigma}_{i}X^{i}_{\lambda}\big)\varphi^{\lambda}\brY_{\bri}^{\nu}\,,
\ea
\label{useful2}
\ee
as well as some projection properties,
\be
\ba{rll}
H^{\mu\rho}H^{\nu\sigma}\whf_{\rho\sigma}&=&(HK)^{\mu}{}_{\rho}(HK)^{\nu}{}_{\sigma}\wtf^{\rho\sigma}\,,\\
H^{\mu\rho}H^{\nu\sigma}u_{\rho\sigma}&=&
(HK)^{\mu}{}_{\rho}(HK)^{\nu}{}_{\sigma}\wtu^{\rho\sigma}\,,\\
~~H^{\mu\rho}u_{\rho\sigma}Y_{i}^{\sigma}&=&(HK)^{\mu}{}_{\rho}\cD_{i}\varphi^{\rho}\,,\\ ~~H^{\mu\rho}u_{\rho\sigma}\brY_{\bri}^{\sigma}&=&(HK)^{\mu}{}_{\rho}\brcD_{\bri}\varphi^{\rho}\,,
\ea
\ee
and
\be
\ba{ll}
K_{\mu\rho}\wtu^{\rho\sigma}X^{i}_{\sigma}=K_{\mu\rho}\cD^{\rho}\varphi^{i}\,,\quad&\quad K_{\mu\rho}\wtu^{\rho\sigma}\brX^{\bri}_{\sigma}=K_{\mu\rho}\brcD^{\rho}\varphi^{\bri}\,,\\
\Upsilon_{\mu}{}^{\nu}K_{\nu\rho}=0=\bar{\Upsilon}_{\mu}{}^{\nu}K_{\nu\rho}\,,\quad&\quad
Y_{i}^{\mu}u_{\mu\nu}\brY^{\nu}_{\bri}=0=X^{i}_{\mu}\wtu^{\mu\nu}\brX^{\bri}_{\nu}\,.
\ea
\ee

Upon taking the flat-spacetime limit~(\ref{flatspacelimit}) with vanishing $B$-field, the connection $\Omega$ vanishes and the covariant derivatives $\fD,\cD$ all reduce to the Yang--Mills covariant derivative, $D = \partial - i[A,~]$. Thus, the field strength~(\ref{FYMcurvedcomponents}),~(\ref{SHN}) and the strain tensor~(\ref{strain}) simplify to~(\ref{FYMcovariant}) and~(\ref{strainflat}), respectively, while from the Lagrangian~(\ref{Fsquare}) we recover~(\ref{sYM}).

Finally, as a curved spacetime generalisation of (\ref{intermediate}), we present the full particle action minimally coupled to the doubled Maxwell vector potential on a generic  $(n,\brn)$ curved background,
\be
\ba{lll}
\cL_{q}&=&{\frac{1}{2}}e^{-1}\rmD_{\tau}x^{A}\rmD_{\tau}x^{B}\cH_{AB}-\half em^{2}-q\rmD_{\tau}x^{A}\cV_{A}\\
&=&{\frac{1}{2}}e^{-1}\dot{x}^{\mu}\dot{x}^{\nu}K_{\mu\nu}-\half e\left(m^{2}+q^{2}\varphi^{\mu}\varphi^{\nu}K_{\mu\nu}\right)-q\dot{x}^{\mu}A_{\mu}\\
{}&{}&+{\frac{1}{2}}e^{-1}
\left(\dot{\tx}_{\mu}-\mathbf{a}_{\mu}-B_{\mu\kappa}\dot{x}^{\kappa}
-eqK_{\mu\rho}\varphi^{\rho}\right)H^{\mu\nu}\\
&{}&\qquad\,\times
\left(\dot{\tx}_{\nu}-\mathbf{a}_{\nu}-B_{\nu\lambda}\dot{x}^{\lambda}-eqK_{\nu\sigma}\varphi^{\sigma}
\right)\\
{}&{}&+X^{i}_{\mu}\big(e^{-1}\dot{x}^{\mu}-q\varphi^{\mu}\big)\left(
\dot{\tx}_{\nu}-\mathbf{a}_{\nu}-B_{\nu\rho}\dot{x}^{\rho}
\right)Y_{i}^{\nu}\\
{}&{}&-\brX^{\bri}_{\mu}\big(e^{-1}\dot{x}^{\mu}+q\varphi^{\mu}\big)\left(
\dot{\tx}_{\nu}-\mathbf{a}_{\nu}-B_{\nu\rho}\dot{x}^{\rho}\right)\brY_{\bri}^{\nu}\,.
\ea
\label{DHD0}
\ee
The first line on the right-hand side of the second equality is essentially the usual (undoubled) action for  a charged `relativistic' particle in the Riemannian subspace, with effective mass generated by the displacement vector field,
\be
m_{\rm{eff.}}^{2}=m^{2}+q^{2}\varphi^{\mu}\varphi^{\nu}K_{\mu\nu}\,.
\label{effm}
\ee
The second and third lines are quadratic in the auxiliary variable $\mathbf{a}_{\mu}$ on the Riemannian subspace and hence are to be negligibly  integrated out, resulting in
\be
H^{\mu\nu}\big(\dot{\tx}_{\nu}-\mathbf{a}_{\nu}-B_{\nu\rho}\dot{x}^{\rho}-eqK_{\nu\rho}\varphi^{\rho}\big)=0\,.
\ee
The last two lines are linear in $\mathbf{a}_{\mu}$ and  hence impose  constraints, 
\be
\ba{ll}
X^{i}_{\mu}\left(e^{-1}\dot{x}^{\mu}-q\varphi^{\mu}\right)=0\,,\qquad&\qquad
\brX^{\bri}_{\mu}\left(e^{-1}\dot{x}^{\mu}+q\varphi^{\mu}\right)=0\,,
\ea
\label{immobility}
\ee
which can be unified into a single expression,
\be
e^{-1}\dot{x}^{\rho}Z_{\rho}{}^{\mu}=q\varphi^{\rho}(1-KH)_{\rho}{}^{\mu}\,.
\label{single}
\ee
It is worthwhile to note the conjugate momenta of $x^{\mu}$,
\be
\ba{l}
\!p_{\mu}=e^{-1\!}\big(K_{\mu\nu}{+B_{\mu\rho}}Z_{\nu}{}^{\rho}\big)\dot{x}^{\nu}
{-qA_{\mu}}{-qB_{\mu\nu\!}\big(}Y^{\nu}_{i}X^{i}_{\rho}{+\brY^{\nu}_{\bri}}\brX^{\bri}_{\rho}\big)\varphi^{\rho}\\
~+e^{-1}\big(B_{\mu\rho}H^{\rho\nu}+Z_{\mu}{}^{\nu}\big)\big(\dot{\tx}_{\nu}-\mathbf{a}_{\nu}-B_{\nu\sigma}\dot{x}^{\sigma}-eqK_{\nu\sigma}\varphi^{\sigma}\big)\,,
\ea
\ee
and  some on-shell values for  the tilde coordinates, 
\be
e^{-1}\big(\rmD_{\tau}\tx_{\mu}-B_{\mu\nu}\dot{x}^{\nu}\big)
=Z_{\mu}{}^{\nu}(p_{\nu}+qA_{\nu})+qK_{\mu\nu}\varphi^{\nu}\,,
\ee
where actually only the momenta along the non-Riemannian directions, $p_{i},\brp_{\bri}$, are relevant.

Clearly, (\ref{immobility}) generalises the  saturation velocity~(\ref{velocitysaturation})  as well as the immobility constraint~(\ref{immobility0})   of the constant background~(\ref{flatH}) to the case of a curved background.  Specifically, for a neutral particle of ${q=0}$,  from (\ref{immobility}) we obtain  the vanishing of the velocity along the  (curved non-Riemannian) $X^{i}_{\mu}$ and $\brX^{\bri}_{\nu}$ directions, 
\be
\ba{ll}
X^{i}_{\mu}\dot{x}^{\mu}=0\,,\qquad&\quad
\brX^{\bri}_{\mu}\dot{x}^{\mu}=0\,.
\ea
\ee
Taking the $\tau$-derivative of these gives expressions that may be viewed as  `non-Riemannian geodesic equations',
\be
\ba{ll}
X^{i}_{\mu}\ddot{x}^{\mu}+\partial_{(\mu}X^{i}_{\nu)}\dot{x}^{\mu}\dot{x}^{\nu}=0\,,\qquad&\quad
\brX^{\bri}_{\mu}\ddot{x}^{\mu}+\partial_{(\mu}\brX^{\bri}_{\nu)}\dot{x}^{\mu}\dot{x}^{\nu}=0\,,
\ea
\ee
For genuine curved backgrounds where  $\partial_{(\mu}X^{i}_{\nu)}$ or $\partial_{(\mu}\brX^{\bri}_{\nu)}$ are nontrivial,  this indicates that the immobility of a fracton  is rather non-trivial.  
Related to  this, it is worthwhile to note that the first curved non-Riemannian DFT background reported in \cite{Lee:2013hma} was shown in \cite{Blair:2020gng} to admit only a finite number of  isometries, implying the absence of higher-moment conservations.   Further investigation with more examples is desired. \\

\section{Discussion\label{SECD}}
Existing  field theoretical  intuition on fractons  is largely  based on  dipole conservation for  charged particles. Contradistinctly, in our  scheme the immobility is universal  regardless of charge, since it  originates  from the `geodesic'  particle  action~(\ref{particle}). Accordingly, our  current~(\ref{vJ}) contains  the energy-momentum tensor rather than  a  charge density of any sort.

Analysis on a spinor field is also possible, following   
\cite{Jeon:2011vx,Angus:2018mep}. We merely comment that,  since  DFT vielbeins square to projectors like $V_{A}{}^{p}V_{Bp}=P_{AB}$,  on the  flat  background~(\ref{flatH}) the  doubled Dirac equation   
\be
\slashed{\gamma}\psi=V^{A}{}_{p}\gamma^{p}\partial_{A}\psi=0\,
\ee
gives
\[
0=(\slashed{\gamma})^{2}\psi=P^{AB}\partial_{A}\partial_{B}\psi=\half\cH^{AB}\partial_{A}\partial_{B}\psi
=\half\partial_{a}\partial^{a}\psi\,.
\]
Thus, the massless spinor is also fractonic, like  (\ref{fractonic}).   Dualisation of the full strain-Maxwell model including the non-Riemannian sector~(\ref{strainMaxwell})   and  the connection to D-branes, following \cite{Pretko:2017kvd, Gromov:2017vir, Pretko:2018tit, fractongauge, Pretko:2019omh, Fruchart:2019qnn, Nguyen:2020yve, Choi:2021kmx, Hirono:2021lmd, Cvetkovic, Beekman} and \cite{Geng:2021cmq}, also  deserve  further study. 

The charged particle action~(\ref{SFracton2}), (\ref{DHD0})  is of interest even upon a genuine Riemannian/Minkowskian  $(0,0)$ flat background,
\be
S^{(0,0)}_{q}=\dis{\int}\rd\tau~
\half e^{-1}\dot{x}^{a}\dot{x}_{a}  -\half e(m^{2}+q^{2}\varphi^{a}\varphi_{a}) -q\dot{x}^{a}A_{a}\,.
\label{00Sq}
\ee
Minimally coupled to the doubled vector potential~$\cV_{A}$,  this particle action  naturally  interacts with  the Maxwell vector potential  of photons~$A_{a}$, and further with the elasticity displacement vector of phonons~$\varphi^{b}$, satisfying,  from (\ref{MaxwellEQ}), (\ref{strainEQ}), 
(\ref{Jmu}),  and (\ref{pseudoJ}),
\be
\ba{ll}
\partial_{c}f^{ca}=J^{a}\,,\quad&\quad
J^{a}(x)=\dis{\sum_{\alpha}\int\rmd\tau}~q\dot{x}^{a}_{\alpha}(\tau)\delta^{D\!}\big(x-x_{\alpha}(\tau)\big)\,,\\
\partial_{c}u^{c}{}_{a}=\tJ_{a}\,, \quad&\quad
\tJ_{a}(x)=\dis{\sum_{\alpha}\int\rmd\tau}~eq^{2}\varphi_{a\,}\delta^{D\!}\big(x-x_{\alpha}(\tau)\big)\,.
\ea
\label{00MS}
\ee
This set of equations may  provide a novel  effective description of polarons~\cite{landau1,landau2,Frohlich,Feynmanpolaron}.  Deformations of  a periodic potential of a crystal lattice are described by phonons, or the  displacements of atoms from their equilibrium positions.  Electrons moving inside the crystal interact with the displacements, which is known as electron-phonon coupling.  Such  electrons with the accompanying deformation are called polarons. They  move freely across the crystal, but with increased effective mass.  This polaron picture essentially agrees with (\ref{00Sq}) and (\ref{00MS}) above.  The charged particles can correspond to both  atomic nuclei  and electrons. From (\ref{00MS}), the lattice structure of the atomic nuclei naturally sets  the dual pseudo-current $\tJ_{a}$ and also the strain tensor $u^{ab}$ to be discretely crystallised on the lattice, while the electrons acquire an effective mass~(\ref{00Sq}) from the condensation of the displacement vector $\varphi^{a}$.    
 We recall the effective mass formula~(\ref{effMASS}) and expand the square root, 
\be
\ba{lll}
m_{{\rm{eff.}}}&=&\sqrt{m^{2}+q^{2}\varphi^{a}\varphi_{a}}\\
{}&=&m
\Big[1+\frac{1}{2}\big(\frac{q}{m}\big)^{2}\varphi^{a}\varphi_{a}
-\frac{1}{8}\big(\frac{q}{m}\big)^{4}\big(\varphi^{a}\varphi_{a}\big)^{2}+\cdots\Big]\,.
\ea
\label{meffexpand}
\ee
 We compare this with a well-known formula for the effective mass of a polaron obtained from estimating its self-energy~\cite{Roseler,Larsen},
 \be
 m_{\mathrm{known}}\simeq m\Big[1+\textstyle{\frac{1}{6}}\alpha_{\mathrm{e-ph}}+0.0236(\alpha_{\mathrm{e-ph}})^{2}\Big]\,,
\label{mknown}
\ee 
where $\alpha_{\mathrm{e-ph}}$ is a  dimensionless     electron-phonon  coupling constant.  From the leading-order terms in the two formulae,  we identify $(\frac{q}{m})^{2}\varphi^{a}\varphi_{a}=\textstyle{\frac{1}{3}}\alpha_{\mathrm{e-ph}}$ which in turn gives, from $(\ref{meffexpand})$, $m_{{\rm{eff.}}}/m\simeq 1+\frac{1}{6}\alpha_{\mathrm{e-ph}}-\frac{1}{72}(\alpha_{\mathrm{e-ph}})^{2}$, and hence differs from (\ref{mknown}) at sub-leading order.  We call for  experimental verification.

For the Riemannian subspace  
we have mostly  envisaged  a Minkowskian signature~(\ref{flatH}), such that time can flow without suffering from immobility and that the effective mass~(\ref{effMASS}) is not necessarily bigger than the true mass.  Intriguingly then,  in the case of  time crystals~\cite{time_crystal}, where  
$\varphi^{a}\varphi_{a}$ would be time-like or negative, our formula seems to  predict that the effective mass of polarons in  time crystals should become smaller.   Note that  a time crystal is  a   quantum system of particles for which the ground state is characterised by repetitive periodic motion of the    particles.    


On the other hand,   if we were to choose   the   Euclidean signature $\eta_{ab}=\delta_{ab}$ for the Riemannian subspace 
and let $(n,\brn)=(1,1)$, thereby including two non-Riemannian directions, one temporal and the other spatial~\cite{Ko:2015rha,Park:2016sbw,Morand:2017fnv,Berman:2019izh,Blair:2019qwi,
Cho:2019ofr,Gallegos:2020egk,Blair:2020gng,Blair:2021ycc}, the corresponding fracton physics would match  that of the 
non-relativistic string~\cite{Gomis:2000bd,Danielsson:2000gi,Gomis:2005pg} and Newton--Cartan 
  gravities~\cite{Christensen:2013lma,Hartong:2015zia,Harmark:2017rpg,
Harmark:2018cdl,Bergshoeff:2018yvt,Bergshoeff:2019pij,
Harmark:2019upf,Bergshoeff:2021bmc}.  Eq.~(\ref{velocitysaturation}) then further implies that time therein can start to flow if the timelike displacement  vector condenses, setting
\be
\dot{t}=eq\varphi^{t}\,.
\ee

It would be of utmost interest if any of the non-Riemannian  geometries underlying  the modified Maxwell equations~(\ref{MaxwellEQ}),  (\ref{strainEQ}) are realised in  Nature.   
 Some well-known singularities in GR~\cite{Witten:1991yr,Burgess:1994kq,Horowitz:1991cd}  
have   recently been   identified  as regular non-Riemannian geometries~\cite{Morand:2021xeq}.  Approaching them,  geodesics indeed become immobile. 
Extra dimensions, if any, might be non-Riemannian~\cite{Morand:2017fnv} and therefore  fractonic. \\

~\\
\section*{Acknowledgments} 
We wish to thank Chris Blair,  Yuji Hirono, Ki-Seok Kim, and Kevin Morand for helpful discussions.  This work is   supported by  the National Research Foundation of Korea  Grants funded by the Korea government~(MSIT),  NRF-2016R1D1A1B01015196, NRF-2018H1D3A1A01030137 (Brain Pool Program),  NRF-2020R1A6A1A03047877 (Center for Quantum Space Time),  NRF-2022R1I1A1A01069032, and NRF-2022R1F1A1070999.  
SA also acknowledges support from the JRG Program at the APCTP through the Science and Technology Promotion Fund and Lottery Fund of the Korean Government, and further  from the Korean Local Governments of Gyeongsangbuk-do Province and Pohang City.  
\\

\hfill


\appendix

\onecolumngrid
\begin{center}
\Large{\bf{Appendix}}\hfill
\end{center}
\section{Doubled Yang--Mills Energy-Momentum Tensor on a Constant Flat Background\label{SECA}}
In this Appendix, we write down the explicit components of the energy-momentum tensor~(\ref{YMT}) for the doubled Yang--Mills theory~(\ref{YM}).  In terms of undoubled spacetime indices, the relevant pieces appearing in the on-shell conserved current~(\ref{vJ}) are
\be
\ba{l}
T^{\mu\nu}=\Tr\!\Big[
(\cF\cH\cF)^{\mu}{}_{\rho}H^{\rho\nu}-H^{\mu\rho}(\cF\cH\cF)_{\rho}{}^{\nu}+(\cF\cH\cF)^{\mu\rho}Z_{\rho}{}^{\nu}-Z_{\rho}{}^{\mu}(\cF\cH\cF)^{\rho\nu}\Big]
-2\partial_{\lambda}\Tr\big[\varphi^{\lambda}\big(P\cF\brP{+\brP\cF P}\big)^{\mu\nu}\Big]\,,\\
T^{\mu}{}_{\nu}=\Tr\!\Big[
(\cF\cH\cF)^{\mu\rho}K_{\rho\nu}-H^{\mu\rho}(\cF\cH\cF)_{\rho\nu}+(\cF\cH\cF)^{\mu}{}_{\rho}Z_{\nu}{}^{\rho}-Z_{\rho}{}^{\mu}(\cF\cH\cF)^{\rho}{}_{\nu}\Big]-2\partial_{\lambda}\Tr\!\Big[
\varphi^{\lambda}\big(P\cF\brP{+\brP\cF P}\big)^{\mu}{}_{\nu}\Big] +\delta^{\mu}{}_{\nu}L_{\YM}\,.
\ea
\label{proceedtoT}
\ee
To evaluate these we need the explicit expressions,
\be
\ba{l}
(\cF\cH\cF)^{\mu\nu}=-D_{\rho}\varphi^{\mu}H^{\rho\sigma}D_{\sigma}\varphi^{\nu}+iZ_{\rho}{}^{\sigma}\big(D_{\sigma}\varphi^{\mu}[\varphi^{\rho},\varphi^{\nu}]-[\varphi^{\mu},\varphi^{\rho}]D_{\sigma}\varphi^{\nu}\big)-[\varphi^{\mu},\varphi^{\rho}]K_{\rho\sigma}[\varphi^{\sigma},\varphi^{\nu}]\,,\\
(\cF\cH\cF)^{\mu}{}_{\nu}=-D_{\rho}\varphi^{\mu}H^{\rho\sigma}f_{\sigma\nu}+i[\varphi^{\mu},\varphi^{\rho}]\big(K_{\rho\sigma}D_{\nu}\varphi^{\sigma}-Z_{\rho}{}^{\sigma}f_{\sigma\nu}\big)+Z_{\rho}{}^{\sigma}D_{\sigma}\varphi^{\mu}D_{\nu}\varphi^{\rho}\,,\\
(\cF\cH\cF)_{\mu}{}^{\nu}=f_{\mu\rho}H^{\rho\sigma}D_{\sigma}\varphi^{\nu}-i\big(D_{\mu}\varphi^{\rho}K_{\rho\sigma}+f_{\mu\rho}Z_{\sigma}{}^{\rho}\big)[\varphi^{\sigma},\varphi^{\nu}]+D_{\mu}\varphi^{\rho}Z_{\rho}{}^{\sigma}D_{\sigma}\varphi^{\nu}\,,\\
(\cF\cH\cF)_{\mu\nu}=f_{\mu\rho}H^{\rho\sigma}f_{\sigma\nu}-D_{\mu}\varphi^{\rho}D_{\nu}\varphi^{\sigma}K_{\rho\sigma}+Z_{\rho}{}^{\sigma}\big(D_{\mu}\varphi^{\rho}f_{\sigma\nu}-f_{\mu\sigma}D_{\nu}\varphi^{\rho}\big)\,,
\ea
\ee
as well as
\be
\ba{l}
-2\big(P\cF\brP{+\brP\cF P}\big)^{\mu\nu}=H^{\mu\rho}H^{\nu\sigma}f_{\rho\sigma}+2H^{\rho[\mu}Z_{\sigma}{}^{\nu]}D_{\rho}\varphi^{\sigma}+i[\varphi^{\mu},\varphi^{\nu}]-i[\varphi^{\rho},\varphi^{\sigma}]Z_{\rho}{}^{\mu}Z_{\sigma}{}^{\nu}\,,
\\
-2\big(P\cF\brP{+\brP\cF P}\big)^{\mu}{}_{\nu}=D_{\nu}\varphi^{\mu}+H^{\mu\rho}D_{\rho}\varphi^{\sigma}K_{\sigma\nu}-i[\varphi^{\rho},\varphi^{\sigma}]Z_{\rho}{}^{\mu}K_{\sigma\nu}+\big(H^{\mu\rho}f_{\rho\sigma}-Z_{\rho}{}^{\mu}D_{\sigma}\varphi^{\rho}\big)Z_{\nu}{}^{\sigma}\,.
\ea
\ee

Substituting these into (\ref{proceedtoT}),  we acquire all the components of the doubled Yang--Mills energy-momentum tensor,
\be
\ba{lll}
T^{a}{}_{b}&=&\Tr\Big[ 
f^{ac}f_{bc}+D^{a}\varphi^{c}D_{b}\varphi_{c}-D_{c}\varphi^{a}D^{c}\varphi_{b}
+[\varphi^{a},\varphi^{c}][\varphi_{b},\varphi_{c}]
+\partial_{\lambda}\big(\varphi^{\lambda}u^{a}{}_{b}\big) 
\\ & & \quad\;\,\, -2Z_{\rho}{}^{\sigma}\Big(f^{(a}{}_{\sigma}D^{c)}\varphi^{\rho}+iD_{\sigma}\varphi^{(a}[\varphi^{c)},\varphi^{\rho}]\Big)\eta_{cb} 
\Big]+\delta^{a}{}_{b}L_{\YM}\,,\\
T^{a}{}_{i}&=&\Tr\Big[
\big(f^{ac}{+D^{c}\varphi^{a}}\big)f_{ic}+D^{-a}\varphi^{c}D_{i}\varphi_{c}+\partial_{\lambda}\big(\varphi^{\lambda}f^{-a}{}_{i}\big)
+Z_{\rho}{}^{\sigma}\big(
f^{-a}{}_{\sigma}D_{i}\varphi^{\rho}+D^{-a}\varphi^{\rho}f_{i\sigma}\big)\Big]\,,
\\
T^{a}{}_{\bri}&=&\Tr\Big[
\big(f^{ac}{-D^{c}\varphi^{a}}\big)f_{\bri c}+D^{+a}\varphi^{c}D_{\bri}\varphi_{c}-\partial_{\lambda}\big(
\varphi^{\lambda}f^{+a}{}_{\bri}\big)
+Z_{\rho}{}^{\sigma}\big(f^{+a}{}_{\sigma}D_{\bri}\varphi^{\rho}+
D^{+a}\varphi^{\rho}f_{\bri\sigma}\big)\Big]\,,\\

T^{i}{}_{a}&=&\Tr\Big[
-D^{c}\varphi^{i}\big(D_{c}\varphi_{a}{+f_{ac}}\big)-i[\varphi^{i},\varphi^{c}]D^{-}_{a}\varphi_{c}+\partial_{\lambda}\big(
\varphi^{\lambda}D^{-}_{a}\varphi^{i}\big)
-Z_{\rho}{}^{\sigma}\big(D_{\sigma}\varphi^{i}D^{-}_{a}\varphi^{\rho}+i[\varphi^{i},\varphi^{\rho}]f^{-}_{a\sigma }\big)\Big]\,,\\

T^{\bri}{}_{a}&=&\Tr\Big[
-D^{c}\varphi^{\bri}\big(D_{c}\varphi_{a}{+f_{c a}}\big)+i[\varphi^{\bri},\varphi^{c}]D^{+}_{a}\varphi_{c}+\partial_{\lambda}\big(
\varphi^{\lambda}D^{+}_{a}\varphi^{\bri}\big)
+Z_{\rho}{}^{\sigma}\big(
D_{\sigma}\varphi^{\bri}D^{+}_{a}\varphi^{\rho}+i[\varphi^{\bri},\varphi^{\rho}]f^{+}_{a\sigma}\big)\Big]\,,

\\
T^{i}{}_{\bri}&=&2\Tr\Big[\,D^{c}\varphi^{i}f_{c\bri}-i[\varphi^{i},\varphi^{c}]D_{\bri}\varphi_{c}+
\partial_{\lambda}\big(\varphi^{\lambda}D_{\bri}\varphi^{i}\big)
-Z_{\rho}{}^{\sigma}\big(D_{\sigma}\varphi^{i}D_{\bri}\varphi^{\rho}+i[\varphi^{i},\varphi^{\rho}]f_{\sigma\bri}\big)\,\Big]\,,
\\
T^{\bri}{}_{i}&=&2\Tr\Big[\,-D^{c}\varphi^{\bri}f_{ci}+i[\varphi^{\bri},\varphi^{c}]D_{i}\varphi_{c}+
\partial_{\lambda}\big(\varphi^{\lambda}D_{i}\varphi^{\bri}\big)
+Z_{\rho}{}^{\sigma}\big(D_{\sigma}\varphi^{\bri}D_{i}\varphi^{\rho}-i[\varphi^{\bri},\varphi^{\rho}]f_{\sigma i}\big)\,\Big]\,, 
\ea
\ee
\be
\ba{lll}
T^{ab}&=&\Tr\!\left[
2D_{c}\varphi^{[a}f^{b]c}-2i[\varphi^{c},\varphi^{[a}]D^{b]}\varphi_{c}+\partial_{\lambda}\Big(
\varphi^{\lambda}\big(f^{ab}+i[\varphi^{a},\varphi^{b}]\big)\Big)
+2Z_{\rho}{}^{\sigma}\big(D_{\sigma}\varphi^{[a}D^{b]}
\varphi^{\rho}-i[\varphi^{\rho},\varphi^{[a}]f^{b]}{}_{\sigma}\big)\right]\,,\\

T^{ai}&=&\Tr\Big[-\big(f^{ac}{+D^{c}\varphi^{a}}\big)D_{c}\varphi^{i}-i[\varphi^{i},\varphi^{c}]D^{-a}\varphi_{c}
+\partial_{\lambda}\big(\varphi^{\lambda}D^{-a}\varphi^{i}\big)
-Z_{\rho}{}^{\sigma}\big(D^{-a}\varphi^{\rho}D_{\sigma}\varphi^{i}-i[\varphi^{\rho},\varphi^{i}]f^{-a}{}_{\sigma}\big)
\Big]\,,\\

T^{a\bri}&=&\Tr\Big[-\big(f^{ac}{-D^{c}\varphi^{a}}\big)D_{c}\varphi^{\bri}-i[\varphi^{\bri},\varphi^{c}]D^{+a}\varphi_{c}
-\partial_{\lambda}\big(\varphi^{\lambda}D^{+a}\varphi^{\bri}\big)
-Z_{\rho}{}^{\sigma}\big(D^{+a}\varphi^{\rho}D_{\sigma}\varphi^{\bri}-i[\varphi^{\rho},\varphi^{\bri}]f^{+a}{}_{\sigma}\big)
\Big]\,,\\
T^{i\bri}&=&2\Tr\Big[\,D^{c}\varphi^{i}D_{c}\varphi^{\bri}-[\varphi^{i},\varphi^{c}][\varphi^{\bri},\varphi_{c}]+i\partial_{\lambda}\big(\varphi^{\lambda}[\varphi^{i},\varphi^{\bri}]\big)+2iZ_{\rho}{}^{\sigma}D_{\sigma}\varphi^{(i}[\varphi^{\bri)},\varphi^{\rho}]\,\Big]\,,
\ea
\label{Tuu}
\ee
and, from the projection properties,
\be
\ba{llll}
T^{ij}=0\,, \qquad&\qquad T^{\bri\brj}=0\,, \qquad&\qquad
T^{i}{}_{j}=\delta^{i}{}_{j}L_{\YM}\,,\qquad&\qquad T^{\bri}{}_{\brj}=\delta^{\bri}{}_{\brj}L_{\YM}\,, \\
T^{ia}= -\eta^{ab}T^{i}{}_{ b}\,, \qquad&\qquad
T^{\bri a}=\eta^{ab}T^{\bri}{}_{ b}\,,
\qquad&\qquad
T^{\bri i}= -T^{i\bri}\,,
\qquad&\qquad
T^{a}{}_{c}\eta^{cb}=T^{b}{}_{c}\eta^{ca}\,.
\ea
\label{vanishingT}
\ee
\twocolumngrid


\begin{thebibliography}{99}


\bibitem{Chamon:2004lew}
C.~Chamon,
``Quantum Glassiness,''
Phys. Rev. Lett. \textbf{94} (2005) no.4, 040402
doi:10.1103/physrevlett.94.040402
[arXiv:cond-mat/0404182 [cond-mat.str-el]].


\bibitem{Haah:2011drr}
J.~Haah,
``Local stabilizer codes in three dimensions without string logical operators,''
Phys. Rev. A \textbf{83} (2011) no.4, 042330
doi:10.1103/physreva.83.042330
[arXiv:1101.1962 [quant-ph]].


\bibitem{Vijay:2015mka}
S.~Vijay, J.~Haah and L.~Fu,
``A New Kind of Topological Quantum Order: A Dimensional Hierarchy of Quasiparticles Built from Stationary Excitations,''
Phys. Rev. B \textbf{92} (2015) no.23, 235136
doi:10.1103/PhysRevB.92.235136
[arXiv:1505.02576 [cond-mat.str-el]].

\bibitem{Vijay:2016phm}
S.~Vijay, J.~Haah and L.~Fu,
``Fracton Topological Order, Generalized Lattice Gauge Theory and Duality,''
Phys. Rev. B \textbf{94} (2016) no.23, 235157
doi:10.1103/PhysRevB.94.235157
[arXiv:1603.04442 [cond-mat.str-el]].





\bibitem{Seiberg:2019vrp}
N.~Seiberg,
``Field Theories With a Vector Global Symmetry,''
SciPost Phys. \textbf{8} (2020) no.4, 050
doi:10.21468/SciPostPhys.8.4.050
[arXiv:1909.10544 [cond-mat.str-el]].


\bibitem{Seiberg:2020bhn}
N.~Seiberg and S.~H.~Shao,
``Exotic Symmetries, Duality, and Fractons in 2+1-Dimensional Quantum Field Theory,''
SciPost Phys. \textbf{10} (2021) no.2, 027
doi:10.21468/SciPostPhys.10.2.027
[arXiv:2003.10466 [cond-mat.str-el]].


\bibitem{Seiberg:2020wsg}
N.~Seiberg and S.~H.~Shao,
``Exotic $U(1)$ Symmetries, Duality, and Fractons in 3+1-Dimensional Quantum Field Theory,''
SciPost Phys. \textbf{9} (2020) no.4, 046
doi:10.21468/SciPostPhys.9.4.046
[arXiv:2004.00015 [cond-mat.str-el]].


\bibitem{Seiberg:2020cxy}
N.~Seiberg and S.~H.~Shao,
``Exotic $\mathbb{Z}_N$ symmetries, duality, and fractons in 3+1-dimensional quantum field theory,''
SciPost Phys. \textbf{10} (2021) no.1, 003
doi:10.21468/SciPostPhys.10.1.003
[arXiv:2004.06115 [cond-mat.str-el]].


\bibitem{Gorantla:2020xap}
P.~Gorantla, H.~T.~Lam, N.~Seiberg and S.~H.~Shao,
``More Exotic Field Theories in 3+1 Dimensions,''
SciPost Phys. \textbf{9} (2020), 073
doi:10.21468/SciPostPhys.9.5.073
[arXiv:2007.04904 [cond-mat.str-el]].


\bibitem{Rudelius:2020kta}
T.~Rudelius, N.~Seiberg and S.~H.~Shao,
``Fractons with Twisted Boundary Conditions and Their Symmetries,''
Phys. Rev. B \textbf{103} (2021) no.19, 195113
doi:10.1103/PhysRevB.103.195113
[arXiv:2012.11592 [cond-mat.str-el]].



\bibitem{Gorantla:2021svj}
P.~Gorantla, H.~T.~Lam, N.~Seiberg and S.~H.~Shao,
``A modified Villain formulation of fractons and other exotic theories,''
J. Math. Phys. \textbf{62} (2021) no.10, 102301
doi:10.1063/5.0060808
[arXiv:2103.01257 [cond-mat.str-el]].


\bibitem{Gorantla:2021bda}
P.~Gorantla, H.~T.~Lam, N.~Seiberg and S.~H.~Shao,
``The low-energy limit of some exotic lattice theories and UV/IR mixing,''
[arXiv:2108.00020 [cond-mat.str-el]].



\bibitem{Nandkishore:2018sel}
R.~M.~Nandkishore and M.~Hermele,
``Fractons,''
Ann. Rev. Condensed Matter Phys. \textbf{10} (2019), 295-313
doi:10.1146/annurev-conmatphys-031218-013604
[arXiv:1803.11196 [cond-mat.str-el]].


\bibitem{Pretko:2020cko}
M.~Pretko, X.~Chen and Y.~You,
``Fracton Phases of Matter,''
Int. J. Mod. Phys. A \textbf{35} (2020) no.06, 2030003
doi:10.1142/S0217751X20300033
[arXiv:2001.01722 [cond-mat.str-el]].




\bibitem{Pretko:2016kxt}
M.~Pretko,
``Subdimensional Particle Structure of Higher Rank U(1) Spin Liquids,''
Phys. Rev. B \textbf{95} (2017) no.11, 115139
doi:10.1103/PhysRevB.95.115139
[arXiv:1604.05329 [cond-mat.str-el]].


\bibitem{Pretko:2016lgv}
M.~Pretko,
``Generalized Electromagnetism of Subdimensional Particles: A Spin Liquid Story,''
Phys. Rev. B \textbf{96} (2017) no.3, 035119
doi:10.1103/PhysRevB.96.035119
[arXiv:1606.08857 [cond-mat.str-el]].

\bibitem{Slagle:2018kqf}
K.~Slagle, A.~Prem and M.~Pretko,
``Symmetric Tensor Gauge Theories on Curved Spaces,''
Annals Phys. \textbf{410} (2019), 167910
doi:10.1016/j.aop.2019.167910
[arXiv:1807.00827 [cond-mat.str-el]].


\bibitem{Gromov:2018nbv}
A.~Gromov,
``Towards classification of Fracton phases: the multipole algebra,''
Phys. Rev. X \textbf{9} (2019) no.3, 031035
doi:10.1103/PhysRevX.9.031035
[arXiv:1812.05104 [cond-mat.str-el]].



\bibitem{Pretko:2017fbf}
M.~Pretko,
``Emergent gravity of fractons: Mach\textquoteright{}s principle revisited,''
Phys. Rev. D \textbf{96} (2017) no.2, 024051
doi:10.1103/PhysRevD.96.024051
[arXiv:1702.07613 [cond-mat.str-el]].



\bibitem{Yan:2018nco}
H.~Yan,
``Hyperbolic fracton model, subsystem symmetry, and holography,''
Phys. Rev. B \textbf{99} (2019) no.15, 155126
doi:10.1103/PhysRevB.99.155126
[arXiv:1807.05942 [hep-th]].




\bibitem{Gromov:2020yoc}
A.~Gromov, A.~Lucas and R.~M.~Nandkishore,
``Fracton hydrodynamics,''
Phys. Rev. Res. \textbf{2} (2020) no.3, 033124
doi:10.1103/PhysRevResearch.2.033124
[arXiv:2003.09429 [cond-mat.str-el]].


\bibitem{Casalbuoni:2021fel}
R.~Casalbuoni, J.~Gomis and D.~Hidalgo,
``World-Line Description of Fractons,''
[arXiv:2107.09010 [hep-th]].





\bibitem{Geng:2021cmq}
H.~Geng, S.~Kachru, A.~Karch, R.~Nally and B.~C.~Rayhaun,
``Fractons and exotic symmetries from branes,''
Fortsch. Phys. \textbf{2021}, 2100133
doi:10.1002/prop.202100133
[arXiv:2108.08322 [hep-th]].



\bibitem{Qi:2020jrf}
M.~Qi, L.~Radzihovsky and M.~Hermele,
``Fracton phases via exotic higher-form symmetry-breaking,''
Annals Phys. \textbf{424} (2021), 168360
doi:10.1016/j.aop.2020.168360
[arXiv:2010.02254 [cond-mat.str-el]].

\bibitem{Distler:2021qzc}
J.~Distler, A.~Karch and A.~Raz,
``Spontaneously Broken Subsystem Symmetries,''
[arXiv:2110.12611 [hep-th]].



\bibitem{Pretko:2017kvd}
M.~Pretko and L.~Radzihovsky,
``Fracton-Elasticity Duality,''
Phys. Rev. Lett. \textbf{120} (2018) no.19, 195301
doi:10.1103/PhysRevLett.120.195301
[arXiv:1711.11044 [cond-mat.str-el]].


\bibitem{Gromov:2017vir}
A.~Gromov,
``Chiral Topological Elasticity and Fracton Order,''
Phys. Rev. Lett. \textbf{122} (2019) no.7, 076403
doi:10.1103/PhysRevLett.122.076403
[arXiv:1712.06600 [cond-mat.str-el]].



\bibitem{Pretko:2018tit}
M.~Pretko and L.~Radzihovsky,
``Symmetry Enriched Fracton Phases from Supersolid Duality,''
Phys. Rev. Lett. \textbf{121} (2018) no.23, 235301
doi:10.1103/PhysRevLett.121.235301
[arXiv:1808.05616 [cond-mat.str-el]].

\bibitem{fractongauge}
 L.~Radzihovsky and M.~Hermele,
``Fractons from Vector Gauge Theory,'
Phys. Rev. Lett. \textbf{124} (2020) 050402 
doi:10.1103/PhysRevLett.124.050402
[arXiv:1905.06951 [cond-mat.str-el].

\bibitem{Pretko:2019omh}
M.~Pretko, Z.~Zhai and L.~Radzihovsky,
``Crystal-to-Fracton Tensor Gauge Theory Dualities,''
Phys. Rev. B \textbf{100} (2019) no.13, 134113
doi:10.1103/PhysRevB.100.134113
[arXiv:1907.12577 [cond-mat.str-el]].

\bibitem{Fruchart:2019qnn}
M.~Fruchart and V.~Vitelli,
``Symmetries and Dualities in the Theory of Elasticity,''
Phys. Rev. Lett. \textbf{124} (2020) no.24, 248001
doi:10.1103/PhysRevLett.124.248001
[arXiv:1912.02384 [cond-mat.soft]].


\bibitem{Nguyen:2020yve}
D.~X.~Nguyen, A.~Gromov and S.~Moroz,
``Fracton-elasticity duality of two-dimensional superfluid vortex crystals: defect interactions and quantum melting,''
SciPost Phys. \textbf{9} (2020), 076
doi:10.21468/SciPostPhys.9.5.076
[arXiv:2005.12317 [cond-mat.quant-gas]].


\bibitem{Choi:2021kmx}
Y.~Choi, C.~Cordova, P.~S.~Hsin, H.~T.~Lam and S.~H.~Shao,
``Non-Invertible Duality Defects in 3+1 Dimensions,''
[arXiv:2111.01139 [hep-th]].


\bibitem{Hirono:2021lmd}
Y.~Hirono and Y.~H.~Qi,
``Effective Field Theories for Gapless Phases with Fractons via a Coset Construction,''
[arXiv:2110.13066 [cond-mat.str-el]].



\bibitem{Bulmash:2018lid}
D.~Bulmash and M.~Barkeshli,
``The Higgs Mechanism in Higher-Rank Symmetric $U(1)$ Gauge Theories,''
Phys. Rev. B \textbf{97} (2018) no.23, 235112
doi:10.1103/PhysRevB.97.235112
[arXiv:1802.10099 [cond-mat.str-el]].


\bibitem{Ma:2018nhd}
H.~Ma, M.~Hermele and X.~Chen,
``Fracton topological order from the Higgs and partial-confinement mechanisms of rank-two gauge theory,''
Phys. Rev. B \textbf{98} (2018) no.3, 035111
doi:10.1103/PhysRevB.98.035111
[arXiv:1802.10108 [cond-mat.str-el]].



\bibitem{Slagle:2020ugk}
K.~Slagle,
``Foliated Quantum Field Theory of Fracton Order,''
Phys. Rev. Lett. \textbf{126} (2021) no.10, 101603
doi:10.1103/PhysRevLett.126.101603
[arXiv:2008.03852 [hep-th]].

\bibitem{Manoj:2020bcz}
N.~Manoj, R.~Moessner and V.~B.~Shenoy,
``Fractonic View of Folding and Tearing Paper: Elasticity of Plates is Dual to a Gauge Theory with Vector Charges,''
Phys. Rev. Lett. \textbf{127} (2021) no.6, 067601
doi:10.1103/PhysRevLett.127.067601
[arXiv:2011.11401 [cond-mat.str-el]].


\bibitem{Radzihovsky:2020xct}
L.~Radzihovsky,
``Quantum smectic gauge theory,''
Phys. Rev. Lett. \textbf{125} (2020) no.26, 267601
doi:10.1103/PhysRevLett.125.267601
[arXiv:2009.06632 [cond-mat.str-el]].



\bibitem{Sun:2021jzs}
K.~Sun and X.~Mao,
``Fractional Excitations in Non-Euclidean Elastic Plates,''
Phys. Rev. Lett. \textbf{127} (2021) no.9, 098001
doi:10.1103/PhysRevLett.127.098001
[arXiv:2101.04186 [cond-mat.soft]].




\bibitem{Yuan:2019geh}
J.~K.~Yuan, S.~A.~Chen and P.~Ye,
``Fractonic Superfluids,''
Phys. Rev. Res. \textbf{2} (2020) no.2, 023267
doi:10.1103/PhysRevResearch.2.023267
[arXiv:1911.02876 [cond-mat.str-el]].

\bibitem{Chen:2020jew}
S.~A.~Chen, J.~K.~Yuan and P.~Ye,
``Fractonic superfluids. II. Condensing subdimensional particles,''
Phys. Rev. Res. \textbf{3} (2021) no.1, 013226
doi:10.1103/PhysRevResearch.3.013226
[arXiv:2010.03261 [cond-mat.str-el]].

\bibitem{Li:2021rga}
H.~Li and P.~Ye,
``Renormalization group analysis on emergence of higher rank symmetry and higher moment conservation,''
Phys. Rev. Res. \textbf{3} (2021) no.4, 043176
doi:10.1103/PhysRevResearch.3.043176
[arXiv:2104.03237 [cond-mat.quant-gas]].


\bibitem{Distler:2021bop}
J.~Distler, M.~Jafry, A.~Karch and A.~Raz,
``Interacting fractons in 2+1-dimensional quantum field theory,''
JHEP \textbf{03} (2022), 070
doi:10.1007/JHEP03(2022)070
[arXiv:2112.05726 [hep-th]].

\bibitem{Yuan:2022mns}
J.~K.~Yuan, S.~A.~Chen and P.~Ye,
``Higher Rank Symmetry Defects: Defect Bound States and Hierarchical Proliferation,''
[arXiv:2201.08597 [cond-mat.str-el]].

\bibitem{Gorantla:2022eem}
P.~Gorantla, H.~T.~Lam, N.~Seiberg and S.~H.~Shao,
``Global Dipole Symmetry, Compact Lifshitz Theory, Tensor Gauge Theory, and Fractons,''
[arXiv:2201.10589 [cond-mat.str-el]].


\bibitem{Yuan:2022wxb}
J.~K.~Yuan, S.~A.~Chen and P.~Ye,
``Quantum Hydrodynamics of Fractonic Superfluids with Lineon Condensate: from Navier-Stokes-like Equations to Landau-like Criterion,''
[arXiv:2203.06984 [cond-mat.supr-con]].

\bibitem{Perez:2022kax}
A.~P\'erez and S.~Prohazka,
``Asymptotic symmetries and soft charges of fractons,''
[arXiv:2203.02817 [hep-th]].



\bibitem{Giveon:1994fu}
A.~Giveon, M.~Porrati and E.~Rabinovici,
``Target space duality in string theory,''
Phys. Rept. \textbf{244} (1994), 77-202
doi:10.1016/0370-1573(94)90070-1
[arXiv:hep-th/9401139 [hep-th]].


\bibitem{Siegel:1993xq}
  W.~Siegel,
  ``Two vierbein formalism for string inspired axionic gravity,''
  Phys.\ Rev.\ D {\bf 47} (1993) 5453
  [hep-th/9302036].

\bibitem{Siegel:1993th}
  W.~Siegel,
  ``Superspace duality in low-energy superstrings,''
  Phys.\ Rev.\ D {\bf 48} (1993) 2826
  [hep-th/9305073].
  
  
\bibitem{Hull:2009mi}
  C.~Hull and B.~Zwiebach,
  ``Double Field Theory,''
  JHEP {\bf 0909} (2009) 099
  [arXiv:0904.4664 [hep-th]].


\bibitem{Hull:2009zb}
  C.~Hull and B.~Zwiebach,
  JHEP {\bf 0909} (2009) 090
  [arXiv:0908.1792 [hep-th]].
  
  
  
  
\bibitem{Hohm:2010jy}
  O.~Hohm, C.~Hull and B.~Zwiebach,
  ``Background independent action for double field theory,''
  JHEP {\bf 1007} (2010) 016
  [arXiv:1003.5027 [hep-th]].
  
  
  
  


\bibitem{Hohm:2010pp}
O.~Hohm, C.~Hull and B.~Zwiebach,
JHEP \textbf{08} (2010), 008
[arXiv:1006.4823 [hep-th]].



\bibitem{Buscher:1987sk}
  T.~H.~Buscher,
  ``A Symmetry of the String Background Field Equations,''
  Phys.\ Lett.\ B {\bf 194} (1987) 59.



\bibitem{Buscher:1987qj}
  T.~H.~Buscher,
  ``Path Integral Derivation of Quantum Duality in Nonlinear Sigma Models,''
  Phys.\ Lett.\ B {\bf 201} (1988) 466.





\bibitem{Jeon:2011cn}
  I.~Jeon, K.~Lee and J.~H.~Park,
  ``Stringy differential geometry, beyond Riemann,''
  Phys.\ Rev.\ D {\bf 84} (2011) 044022
  doi:10.1103/PhysRevD.84.044022
  [arXiv:1105.6294 [hep-th]].







\bibitem{Park:2015bza}
  J.~H.~Park, S.~J.~Rey, W.~Rim and Y.~Sakatani,
  ``$\ODD$ covariant Noether currents and global charges in double field theory,''
  JHEP {\bf 1511} (2015) 131
  [arXiv:1507.07545 [hep-th]].




\bibitem{Hohm:2011si}
O.~Hohm and B.~Zwiebach,
``On the Riemann Tensor in Double Field Theory,''
JHEP \textbf{05} (2012), 126
doi:10.1007/JHEP05(2012)126
[arXiv:1112.5296 [hep-th]].


\bibitem{Angus:2018mep}
  S.~Angus, K.~Cho and J.~H.~Park,
  ``Einstein Double Field Equations,''
  Eur.\ Phys.\ J.\ C {\bf 78} (2018) no.6,  500
  doi:10.1140/epjc/s10052-018-5982-y
  [arXiv:1804.00964 [hep-th]].


\bibitem{Jeon:2011sq}
I.~Jeon, K.~Lee and J.~H.~Park,
``Supersymmetric Double Field Theory: Stringy Reformulation of Supergravity,''
Phys. Rev. D \textbf{85} (2012),  081501(R)  
doi:10.1103/PhysRevD.85.081501
[arXiv:1112.0069 [hep-th]].


\bibitem{Jeon:2012hp}
I.~Jeon, K.~Lee, J.~H.~Park and Y.~Suh,
``Stringy Unification of Type IIA and IIB Supergravities under N=2 D=10 Supersymmetric Double Field Theory,''
Phys. Lett. B \textbf{723} (2013), 245-250
doi:10.1016/j.physletb.2013.05.016
[arXiv:1210.5078 [hep-th]].


\bibitem{Jeon:2011vx}
I.~Jeon, K.~Lee and J.~H.~Park,
``Incorporation of fermions into double field theory,''
JHEP \textbf{11} (2011), 025
doi:10.1007/JHEP11(2011)025
[arXiv:1109.2035 [hep-th]].


\bibitem{Jeon:2011kp}
I.~Jeon, K.~Lee and J.~H.~Park,
``Double field formulation of Yang-Mills theory,''
Phys. Lett. B \textbf{701} (2011), 260-264
doi:10.1016/j.physletb.2011.05.051
[arXiv:1102.0419 [hep-th]].


\bibitem{Jeon:2012kd}
I.~Jeon, K.~Lee and J.~H.~Park,
``Ramond-Ramond Cohomology and O(D,D) T-duality,''
JHEP \textbf{09} (2012), 079
doi:10.1007/JHEP09(2012)079
[arXiv:1206.3478 [hep-th]].


\bibitem{Hohm:2011zr}
O.~Hohm, S.~K.~Kwak and B.~Zwiebach,
``Unification of Type II Strings and T-duality,''
Phys. Rev. Lett. \textbf{107} (2011), 171603
doi:10.1103/PhysRevLett.107.171603
[arXiv:1106.5452 [hep-th]].




\bibitem{Hohm:2011dv}
O.~Hohm, S.~K.~Kwak and B.~Zwiebach,
``Double Field Theory of Type II Strings,''
JHEP \textbf{09} (2011), 013
doi:10.1007/JHEP09(2011)013
[arXiv:1107.0008 [hep-th]].





\bibitem{Angus:2019bqs}
S.~Angus, K.~Cho, G.~Franzmann, S.~Mukohyama and J.~H.~Park,
``$\mathbf {O}(D,D)$ completion of the Friedmann equations,''
Eur. Phys. J. C \textbf{80} (2020) no.9, 830
doi:10.1140/epjc/s10052-020-8379-7
[arXiv:1905.03620 [hep-th]].

\bibitem{Lescano:2021nju}
E.~Lescano and N.~Mir\'on-Granese,
``Double Field Theory with matter and its cosmological application,''
[arXiv:2111.03682 [hep-th]].





\bibitem{Ko:2016dxa}
S.~M.~Ko, J.~H.~Park and M.~Suh,
``The rotation curve of a point particle in stringy gravity,''
JCAP \textbf{06} (2017), 002
doi:10.1088/1475-7516/2017/06/002
[arXiv:1606.09307 [hep-th]].

\bibitem{Basile:2019pic}
T.~Basile, E.~Joung and J.~H.~Park,
``A note on Faddeev--Popov action for doubled-yet-gauged particle and graded Poisson geometry,''
JHEP \textbf{02} (2020), 022
doi:10.1007/JHEP02(2020)022
[arXiv:1910.13120 [hep-th]].




\bibitem{Lee:2013hma}
  K.~Lee and J.~H.~Park,
  ``Covariant action for a string in doubled yet gauged  spacetime,''
  Nucl.\ Phys.\ B {\bf 880} (2014) 134 
  [arXiv:1307.8377 [hep-th]].

\bibitem{Park:2016sbw}
J.~H.~Park,
``Green-Schwarz superstring on doubled-yet-gauged spacetime,''
JHEP \textbf{11} (2016), 005
doi:10.1007/JHEP11(2016)005
[arXiv:1609.04265 [hep-th]].





\bibitem{Choi:2015bga}
K.~S.~Choi and J.~H.~Park,
``Standard Model as a Double Field Theory,''
Phys. Rev. Lett. \textbf{115} (2015) no.17, 171603
doi:10.1103/PhysRevLett.115.171603
[arXiv:1506.05277 [hep-th]].






  

\bibitem{Morand:2017fnv}
  K.~Morand and J.~H.~Park,
  ``Classification of non-Riemannian doubled-yet-gauged spacetime,''
  Eur.\ Phys.\ J.\ C {\bf 77} (2017) no.10,  685
  [arXiv:1707.03713 [hep-th]].



\bibitem{Cho:2018alk}
K.~Cho, K.~Morand and J.~H.~Park,
``Kaluza\textendash{}Klein reduction on a maximally non-Riemannian space is moduli-free,''
Phys. Lett. B \textbf{793} (2019), 65-69
doi:10.1016/j.physletb.2019.04.042
[arXiv:1808.10605 [hep-th]].


\bibitem{Berman:2019izh}
D.~S.~Berman, C.~D.~A.~Blair and R.~Otsuki,
``Non-Riemannian geometry of M-theory,''
JHEP \textbf{07} (2019), 175
doi:10.1007/JHEP07(2019)175
[arXiv:1902.01867 [hep-th]].


\bibitem{Ko:2015rha}
S.~M.~Ko, C.~Melby-Thompson, R.~Meyer and J.~H.~Park,
``Dynamics of Perturbations in Double Field Theory \& Non-Relativistic String Theory,''
JHEP \textbf{12} (2015), 144
doi:10.1007/JHEP12(2015)144
[arXiv:1508.01121 [hep-th]].


\bibitem{Blair:2019qwi}
C.~D.~A.~Blair,
``A worldsheet supersymmetric Newton-Cartan string,''
JHEP \textbf{10} (2019), 266
doi:10.1007/JHEP10(2019)266
[arXiv:1908.00074 [hep-th]].


\bibitem{Gomis:2000bd}
  J.~Gomis and H.~Ooguri,
  ``Nonrelativistic closed string theory,''
  J.\ Math.\ Phys.\  {\bf 42} (2001) 3127
  doi:10.1063/1.1372697
  [hep-th/0009181].

\bibitem{Danielsson:2000gi}
U.~H.~Danielsson, A.~Guijosa and M.~Kruczenski,
``IIA/B, wound and wrapped,''
JHEP \textbf{10} (2000), 020
doi:10.1088/1126-6708/2000/10/020
[arXiv:hep-th/0009182 [hep-th]].

\bibitem{Andringa:2012uz}
R.~Andringa, E.~Bergshoeff, J.~Gomis and M.~de Roo,
``'Stringy' Newton-Cartan Gravity,''
Class. Quant. Grav. \textbf{29} (2012), 235020
doi:10.1088/0264-9381/29/23/235020
[arXiv:1206.5176 [hep-th]].


\bibitem{Bergshoeff:2014jla}
E.~Bergshoeff, J.~Gomis and G.~Longhi,
``Dynamics of Carroll Particles,''
Class. Quant. Grav. \textbf{31} (2014) no.20, 205009
doi:10.1088/0264-9381/31/20/205009
[arXiv:1405.2264 [hep-th]].

\bibitem{Duval:2014uoa}
C.~Duval, G.~W.~Gibbons, P.~A.~Horvathy and P.~M.~Zhang,
``Carroll versus Newton and Galilei: two dual non-Einsteinian concepts of time,''
Class. Quant. Grav. \textbf{31} (2014), 085016
doi:10.1088/0264-9381/31/8/085016
[arXiv:1402.0657 [gr-qc]].


\bibitem{Bekaert:2015xua}
X.~Bekaert and K.~Morand,
``Connections and dynamical trajectories in generalised Newton-Cartan gravity II. An ambient perspective,''
J. Math. Phys. \textbf{59} (2018) no.7, 072503
[arXiv:1505.03739 [hep-th]].




\bibitem{Harmark:2017rpg}
T.~Harmark, J.~Hartong and N.~A.~Obers,
``Nonrelativistic strings and limits of the AdS/CFT correspondence,''
Phys. Rev. D \textbf{96} (2017) no.8, 086019
doi:10.1103/PhysRevD.96.086019
[arXiv:1705.03535 [hep-th]].



\bibitem{Park:2013mpa}
J.~H.~Park,
``Comments on double field theory and diffeomorphisms,''
JHEP \textbf{06} (2013), 098
doi:10.1007/JHEP06(2013)098
[arXiv:1304.5946 [hep-th]].

\bibitem{Blair:2020gng}
C.~D.~A.~Blair, G.~Oling and J.~H.~Park,
``Non-Riemannian isometries from double field theory,''
JHEP \textbf{04} (2021), 072
doi:10.1007/JHEP04(2021)072
[arXiv:2012.07766 [hep-th]].






\bibitem{Park:2017snt}
  J.~H.~Park,
  ``Stringy Gravity: Solving the Dark Problems at ‘short’ distance,''
  EPJ Web Conf.\  {\bf 168} (2018) 01010
  doi:10.1051/epjconf/201816801010
  [arXiv:1707.08961 [hep-th]].



\bibitem{Cvetkovic}
V.~Cvetkovic, Z.~Nussinov and J.~Zaanen,
 ``Topological kinematic constraints: quantum dislocations and the glide principle,''
Philosophical Magazine, 86 (2006), 2995. doi:10.1080/14786430600636328


\bibitem{Park:2020ixf}
J.~H.~Park and S.~Sugimoto,
``String Theory and non-Riemannian Geometry,''
Phys. Rev. Lett. \textbf{125} (2020) no.21, 211601
doi:10.1103/PhysRevLett.125.211601
[arXiv:2008.03084 [hep-th]].

  

\bibitem{Jeon:2010rw}
I.~Jeon, K.~Lee and J.~H.~Park,
``Differential geometry with a projection: Application to double field theory,''
JHEP \textbf{04} (2011), 014
doi:10.1007/JHEP04(2011)014
[arXiv:1011.1324 [hep-th]].
  



  
\bibitem{Guddala2021}
 S.~Guddala, F.~Komissarenko, S.~Kiriushechkina, A.~Vakulenko, M.~Li, V.~M.~Menon,   A.~Al\`{u}, A.~B.~Khanikaev, 
 ``Topological phonon-polariton funneling in midinfrared metasurfaces,'' 
Science. 2021 Oct 8;374(6564):225-227. 
doi: 10.1126/science.abj5488. Epub 2021 Oct 7. PMID: 34618590.





\bibitem{Cho:2019ofr}
K.~Cho and J.~H.~Park,
``Remarks on the non-Riemannian sector in Double Field Theory,''
Eur. Phys. J. C \textbf{80} (2020) no.2, 101
doi:10.1140/epjc/s10052-020-7648-9
[arXiv:1909.10711 [hep-th]].




\bibitem{Bidussi:2021nmp}
L.~Bidussi, J.~Hartong, E.~Have, J.~Musaeus and S.~Prohazka,
``Fractons, dipole symmetries and curved spacetime,''
[arXiv:2111.03668 [hep-th]].

\bibitem{Jain:2021ibh}
A.~Jain and K.~Jensen,
``Fractons in curved space,''
[arXiv:2111.03973 [hep-th]].



\bibitem{Milne:1934}
E .A. Milne, ``A Newtonian Expanding Universe,"  Quat. J. Math. (Oxford Ser.) 5
(1934), 64-72.



\bibitem{Duval:1993pe}
C.~Duval,
``On Galileian isometries,''
Class. Quant. Grav. \textbf{10} (1993), 2217-2222
doi:10.1088/0264-9381/10/11/006
[arXiv:0903.1641 [math-ph]].


\bibitem{Beekman} 
A. J.~Beekman, J.~ Nissinen, K.~Wu, R.-J~Slager, Z.~Nussinov, V.~Cvetkovic and J.~Zaanen, 
 ``Dual gauge field theory of quantum liquid crystals in two dimensions,''
Phys. Rept. \textbf{683} (2017), 1
doi:10.1016/j.physrep.2017.03.004






\bibitem{landau1}
L.~D. ~Landau,  
``\"Uber die Bewegung der Elektronen in Kristallgitter,'' 
Phys. Z. Sowjetunion   3 (1933), 644–645.

\bibitem{landau2}
L.~D. ~Landau and S.~.I. ~Pekar,  
``Effective mass of a polaron,'' 
Zh. Eksp. Teor. Fiz. 18 (1948), 419–423

\bibitem{Frohlich}
H.~ Fr\"ohlich H, 
``Electrons in lattice fields,''  
Adv. Phys. 3 (1954) (11)
doi:10.1080/00018735400101213

\bibitem{Feynmanpolaron}
R.~P. ~Feynman,
``Slow Electrons in a Polar Crysta,'' 
Phys. Rev. \textbf{97} (1955) no.3, 660
doi:10.1103/PhysRev.97.660.


\bibitem{Roseler}
J. R$\rm{\ddot{o}}$seler, 
``A new variational ansatz in the polaron theory,''
Physica Status Solidi B. \textbf{25} (1968) 311 
doi:10.1002/pssb.19680250129

\bibitem{Larsen}
D. M. Larsen, 
``Intermediate-Coupling Polaron Effective Mass,'' 
Phys. Rev. \textbf{174} (1968) 1046 
doi:10.1103/PhysRev.174.1046





\bibitem{time_crystal}
 A. D. Shapere and F. Wilczek, 
 ``Regularizations of time-crystal dynamics,"  
 PNAS September 17, 2019 116 (38) 18772-18776 
 doi:10.1073/pnas.1908758116  





\bibitem{Gallegos:2020egk}
A.~D.~Gallegos, U.~G\"ursoy, S.~Verma and N.~Zinnato,
``Non-Riemannian gravity actions from double field theory,''
[arXiv:2012.07765 [hep-th]].




\bibitem{Blair:2021ycc}
C.~D.~A.~Blair, D.~Gallegos and N.~Zinnato,
``A non-relativistic limit of M-theory and 11-dimensional membrane Newton-Cartan,''
[arXiv:2104.07579 [hep-th]].







\bibitem{Gomis:2005pg}
J.~Gomis, J.~Gomis and K.~Kamimura,
``Non-relativistic superstrings: A New soluble sector of $AdS(5) \times S^5$,''
JHEP \textbf{12} (2005), 024
doi:10.1088/1126-6708/2005/12/024
[arXiv:hep-th/0507036 [hep-th]].



\bibitem{Christensen:2013lma}
M.~H.~Christensen, J.~Hartong, N.~A.~Obers and B.~Rollier,
``Torsional Newton-Cartan Geometry and Lifshitz Holography,''
Phys. Rev. D \textbf{89} (2014), 061901(R)
doi:10.1103/PhysRevD.89.061901
[arXiv:1311.4794 [hep-th]].

\bibitem{Hartong:2015zia}
J.~Hartong and N.~A.~Obers,
``Ho\v{r}ava-Lifshitz gravity from dynamical Newton-Cartan geometry,''
JHEP \textbf{07} (2015), 155
doi:10.1007/JHEP07(2015)155
[arXiv:1504.07461 [hep-th]].



\bibitem{Harmark:2018cdl}
T.~Harmark, J.~Hartong, L.~Menculini, N.~A.~Obers and Z.~Yan,
``Strings with Non-Relativistic Conformal Symmetry and Limits of the AdS/CFT Correspondence,''
JHEP \textbf{11} (2018), 190
doi:10.1007/JHEP11(2018)190
[arXiv:1810.05560 [hep-th]].

\bibitem{Bergshoeff:2018yvt}
E.~Bergshoeff, J.~Gomis and Z.~Yan,
``Nonrelativistic String Theory and T-Duality,''
JHEP \textbf{11} (2018), 133
doi:10.1007/JHEP11(2018)133
[arXiv:1806.06071 [hep-th]].

\bibitem{Bergshoeff:2019pij}
E.~A.~Bergshoeff, J.~Gomis, J.~Rosseel, C.~\c{S}im\c{s}ek and Z.~Yan,
``String Theory and String Newton-Cartan Geometry,''
J. Phys. A \textbf{53} (2020) no.1, 014001
doi:10.1088/1751-8121/ab56e9
[arXiv:1907.10668 [hep-th]].

\bibitem{Harmark:2019upf}
T.~Harmark, J.~Hartong, L.~Menculini, N.~A.~Obers and G.~Oling,
``Relating non-relativistic string theories,''
JHEP \textbf{11} (2019), 071
doi:10.1007/JHEP11(2019)071
[arXiv:1907.01663 [hep-th]].


\bibitem{Bergshoeff:2021bmc}
E.~A.~Bergshoeff, J.~Lahnsteiner, L.~Romano, J.~Rosseel and C.~\c{S}im\c{s}ek,
``A Non-Relativistic Limit of NS-NS Gravity,''
[arXiv:2102.06974 [hep-th]].











\bibitem{Witten:1991yr}
E.~Witten,
``On string theory and black holes,'' 
Phys. Rev. D \textbf{44} (1991), 314-324 
doi:10.1103/PhysRevD.44.314



\bibitem{Burgess:1994kq}
C.~P.~Burgess, R.~C.~Myers and F.~Quevedo,
``On spherically symmetric string solutions in four-dimensions,''
Nucl. Phys. B \textbf{442} (1995), 75-96
doi:10.1016/S0550-3213(95)00090-9
[arXiv:hep-th/9410142 [hep-th]].



\bibitem{Horowitz:1991cd}
G.~T.~Horowitz and A.~Strominger,
``Black strings and P-branes,''
Nucl. Phys. B \textbf{360} (1991), 197-209
doi:10.1016/0550-3213(91)90440-9




\bibitem{Morand:2021xeq}
K.~Morand, J.~H.~Park and M.~Park,
``Identifying Riemannian Singularities with Regular Non-Riemannian Geometry,''
Phys. Rev. Lett. \textbf{128} (2022) no.4, 041602
doi:10.1103/PhysRevLett.128.041602
[arXiv:2106.01758 [hep-th]].





\end{thebibliography}
\end{document}